\documentclass[12pt]{article}
 \usepackage[numbers]{natbib} 

\usepackage[preprint]{nips_2018}

\usepackage[utf8]{inputenc} 
\usepackage[T1]{fontenc}    
\usepackage[draft=false]{hyperref}       
\usepackage{booktabs}       
\usepackage{amsfonts}       
\usepackage{nicefrac}       
\usepackage[draft=false]{microtype}      
\usepackage{graphicx}
\usepackage{todonotes}
\usepackage{mathtools}
\usepackage{color}
\usepackage{pgfplots}
\usepackage{enumitem}
\usepackage{textcomp,gensymb}
\usepackage{subcaption}
\usepackage{array}
\usepackage[export]{adjustbox}
\usepackage{multirow}
\newcommand{\sa}{2.90em}
\newcommand{\s}{2.40em}
\newcommand{\wa}{0.075}
\newcommand{\w}{0.075}
\newcommand{\norm}[1]{\|#1\|}

 \definecolor{amber}{rgb}{1.0, 0.75, 0.0}
 \definecolor{ube}{rgb}{0.53, 0.47, 0.76}
\definecolor{seabornback}{HTML}{EAEAF2}

\pgfplotsset{
    legend image with text/.style={
        legend image code/.code={%
            \node[anchor=center] at (0.3cm,0cm) {#1};
        }
    },
}
\title{Physics-aware Deep Generative Models for \\ Creating Synthetic Microstructures}

\author{
 Rahul Singh\thanks{Equal contribution. This work was supported in part by grants NSF CCF-1750920, CCF-1815101m, a Faculty Fellowship by the Black and Veatch Foundation, and an equipment grant from the NVIDIA Corporation (C.H., R.S., V.S.); NSF 1149365 (B.G., B.S.S.P); U.S. AFOSR YIP grant FA9550-17-1-0220 (S.S.).}, Viraj Shah\footnotemark[1], Balaji Pokuri \\
 Iowa State University\\
  Ames, IA 50010 \\
  \texttt{rahulsn@iastate.edu, viraj@iastate.edu, balajip@iastate.edu} \\
   \And
  Soumik Sarkar, Baskar Ganapathysubramanian, Chinmay Hegde\\ 
   Iowa State University\\
  Ames, IA 50010 \\
  \texttt{soumiks@iastate.edu, baskarg@iastate.edu, chinmay@iastate.edu}
}

\begin{document}

\maketitle

\begin{abstract}
A key problem in computational material science deals with understanding the effect of material distribution (i.e., microstructure) on material performance. The challenge is to synthesize microstructures, given a finite number of microstructure images, and/or some physical invariances that the microstructure exhibits. Conventional approaches are based on stochastic optimization and are computationally intensive. 
We introduce three generative models for the fast synthesis of binary microstructure images. The first model is a WGAN model that uses a finite number of training images to synthesize new  microstructures that weakly satisfy the physical invariances respected by the original data. The second model explicitly enforces known physical invariances by replacing the traditional discriminator in a GAN with an invariance checker. Our third model combines the first two models to reconstruct microstructures that respect both explicit physics invariances as well as implicit constraints learned from the image data. 
We illustrate these models by reconstructing two-phase microstructures that exhibit coarsening behavior. 
The trained models also exhibit interesting latent variable interpolation behavior, and the results indicate considerable promise for enforcing user-defined physics constraints during microstructure synthesis. 

\end{abstract}

\section{Introduction}
\label{sec:intro}
\subsection{Motivation}

An overarching theme of materials research is the design of material distributions (also called microstructure) so that the ensuing material exhibits tailored properties. The final property of the material system is intricately connected to the underlying microstructure. This theme encompasses several material systems including porous materials~\cite{torquato2013random}, steels and welds~\cite{Bhadeshia}, composites~\cite{Li2000}, powder metallurgy~\cite{wang2010critical}, 3D printing~\cite{Schumacher}, energy storage devices as batteries~\cite{Garcia-Garcia2016}, and energy converting devices like bulk hetero-junction solar cells~\cite{Rivnay2012}. Microstructure-sensitive design has been used to tailor a wide variety of properties including strengths, heat and mass diffusivities, energy storage capacity and lifetime, and energy conversion efficiency. 
In microstructure-sensitive design, quantifying the effect of microstructure features on performance is critical for the efficient design of application-tailored devices.

There are extensive efforts to experimentally image the microstructure (including X-ray, optical, and electron microscopy). However, getting a complete virtual instance of a microstructure is non-trivial, as is the process of perturbing specific features of the microstructure. Thus, an entire sub-field in computational material science is devoted to the development of methods for the \emph{simulation} of microstructures~\cite{Ganapathysubramanian2008, Ganapathysubramanian2007, roberts1997statistical}. Here, microstructure realizations are synthesized that satisfy certain target statistical properties of the material distribution. These statistical properties could be scalars (such like total volume fraction of a material) or more complex functions (like 2-point correlations and other material statistics)~\cite{torquato2013random}.


\subsection{Our contributions}

Over the past several decades, a number of methods have been developed for microstructure simulations, including Gaussian random fields~\cite{roberts1997statistical}, optimization-based methods~\cite{torquato1998reconstructing}, multi-point statistics~\cite{Feng2018}, and layer-by-layer reconstruction~\cite{tahmasebi2012multiple}. Most of these techniques use some form of optimization to refine an initial microstructure to satisfy target constraints. They are formulated as a minimization of the difference between the constraints/invariances of the simulated microstructures and the target. Consequently, these approaches are computationally intensive, requiring several compute hours even to simulate just \emph{one} synthetic example. 
Such simulation methods are prohibitively slow to perform analysis for complex problems, which typically involve millions of pixels and require multiple constraints/invariances satisfaction. 

In this paper, we introduce novel techniques that use generative adversarial networks (GANs) to generate microstructures. We propose and test three separate techniques: 

\begin{enumerate}[leftmargin=*]
\item The first approach uses a standard GAN architecture trained with a Wasserstein metric. 
We show that the generated images respect the distribution of certain physical invariances --- specifically, the one-point and two-point correlations --- of the training data. In this approach, we effectively let the discriminator learn the features of the data. 
\item The second approach replaces the traditional discriminator with a \emph{checker} function. This checker function is defined by the user and identifies the most physically informed features of the microstructures. There is no discriminator training involved and thus the issue of mode collapse can be averted. Here, the data requirements are minimal, and data is only used to calibrate the checker. We demonstrate how the generated images are diverse and mimic the supplied invariance metric (two point correlation curve). 
\item Finally, we propose a hybrid of the above two architectures. We demonstrate its potential to simultaneously assimilate patterns both from the available data and user description. This enables the exploration and replication of non-quantifiable phenomena in the data, along with user-defined constraints.
\end{enumerate}

We validate our techniques using a range of numerical experiments. We employ a 2D microstructure dataset that is generated by simulating a phase separation process of two (immiscible) components under thermal annealing~\cite{Wodo2012}. Diverse microstructures can be generated, of varying domain-purity, domain size, interfacial area and (relative) volume fractions of the components. 
Overall, our results indicate that the models that we obtain have considerable promise to not just capture visually salient microstructure features but \emph{also the physics} underlying the data generation.

\section{Proposed techniques}
\label{sec:methods}
In the context of this paper, we consider the underlying material to be a two-phase homogeneous, isotropic material. 
Our setup for statistical characterization of microstructure follows with Torquato \cite{torquato2013random}. Consider an instance of the two-phase homogeneous isotropic material within $d$-dimensional Euclidean space $\mathbb{R}^d$ (where $d \in \{2,3\}$). A phase function $\phi(\cdot)$ is used to characterize this two-phase system, defined as:
\begin{equation}
    \phi^{(1)}(\mathbf{r}) = \begin{cases}
                                1, \mathbf{r} \in V_1,\\
                                0, \mathbf{r} \in V_2,
                        \end{cases}
\end{equation}
where $V_1 \in \mathbb{R}^d$ is the region occupied by phase 1 and $V_2 \in \mathbb{R}^d$ is the region occupied by phase 2. 

Given this microstructure defined by the phase function, $\phi$, statistical characteristics can be evaluated. These include the $n$-point correlation functions for $n ={1,2,3,...}$. For homogeneous and isotropic media, $n-$point correlations depend neither on the absolute positions of $n-$points, nor on the rotation of these spatial co-ordinates; instead, they depend only on relative displacements. The $1$-point correlation function, $p_1$, commonly known as \emph{volume fraction}, is constant throughout the material. The volume fraction of phase 1, $p_1^{(1)}$, is defined as:
$$
p_1^{(1)} = \mathbb{E}_{\mathbf{r}} \phi^{(1)}(\mathbf{r}).
$$
The $2-$point correlation is defined as: 
$$
p_2^{(1)}(r_{12}) = \mathbb{E}_{\mathbf{r_1},\mathbf{r_2}} \left[\phi^{(1)}(\mathbf{r}_1)\phi^{(1)}(\mathbf{r}_2)\right].
$$

The $2-$point correlation is one of the most important statistical descriptors of microstructures. An alternate interpretation of 2-point correlation is the probability that two randomly chosen points $\mathbf{r}_1$ and $\mathbf{r}_2$ a certain distance apart both share the same phase. 

Henceforth we omit the superscript representing the phase and subscripts representing the spatial points for simplicity, and refer to volume fraction as $p_1$, and $2$-point correlation as $p_2$. It can be shown that $p_2(r = 0) = p_1$ and $\lim_{r\to\infty} p_2(r) = p_1^2$.


We now propose three generative models capable of generating grayscale microstructure images of two-phase materials. Each model gives us a varying degree of control over the statistical invariances($p_1$ and $p_2$) of the generated samples. 

\subsection{Wasserstein GAN}

Generative Adversarial Networks (GAN) are powerful models that attempt to learn high-dimensional data distributions such as images \cite{goodfellow2014generative}. The key strategy is to pose the estimation of generative model as two-player minimax game, with each player being a trainable model, referred as the generator and discriminator respectively. Both the networks are trained simultaneously in an adversarial manner. The generator is trained to generate realistic images by learning a nonlinear mapping from a low-dimensional space of latent parameters to the space of real images, while the discriminator is trained to discriminate or classify the samples generated by the generator (G) as either `real' or `fake'. 
After sufficient training, the generator network is (often) able to reproduce synthetic images that closely resemble the original images. Formally, the minimax objective of GAN can be represented as:
\begin{equation}
 \min_{G} \max_{D} \underset{\mathbf{x}\sim \mathbb{P}_r}{\mathbb{E}} \left[\log{\left(D(\mathbf{x})\right)}\right] + \underset{\mathbf{\tilde{x}}\sim \mathbb{P}_g}{\mathbb{E}} \left[\log{\left(1-D(\mathbf{\tilde{x}})\right)}\right],
 \label{eq:gan}
\end{equation}
where $\mathbb{P}_r$ is the real data distribution, and $\mathbb{P}_g$ is data distribution of images generated by generator. $\mathbf{\tilde{x}}$ is defined as $\mathbf{\tilde{x}} = G(\mathbf{z})$ with $\mathbf{z}$ being drawn from noise distribution $p$, typically a Gaussian distribution. If the discriminator is trained optimally, then at each generator update the minimization of the function in Eq.~\ref{eq:gan} turns out to be the minimization of the Jenson-Shannon divergence between $\mathbb{P}_r$ and $\mathbb{P}_g$.

The GAN framework proposed in \cite{goodfellow2014generative} can produce visually appealing samples, but it usually suffers from training instability. The reason for the training difficulty, as explained in \cite{Arjovsky2017WassersteinGA} is that the divergence is not continuous with respect to the  parameters of the generator. An proposed improvement, known as Wasserstein GAN (WGAN), advocates using the Earth Mover Distance (also known as the Wasserstein-1 distance), which is continuous under reasonable assumptions. Moreover, the training can be stabilized via a suitable weight-clipping step in each epoch \cite{Arjovsky2017WassersteinGA}. 

However, even the WGAN can sometimes generate poor samples and fail to converge. To tackle the issue, an alternative to weight clipping is proposed in \cite{gulrajani2017improved}, which stabilizes the GAN training by penalizing the norm of the gradient of the discriminator with respect to its input. The new objective function, which is a combination of Wasserstein distance and a gradient penalty (GP), becomes:
\begin{equation}
L = \underbrace{\underset{\mathbf{\tilde{x}}\sim \mathbb{P}_g}{\mathbb{E}} \left[D(\mathbf{\tilde{x}})\right] - \underset{\mathbf{x}\sim \mathbb{P}_r}{\mathbb{E}} \left[D(\mathbf{x})\right]}_\text{Wasserstein loss}~ + ~\underbrace{\lambda \underset{\mathbf{\tilde{x}}\sim \mathbb{P}_{\mathbf{\tilde{x}}}}{\mathbb{E}} \left[ \left(\norm{\nabla_{\mathbf{\hat{x}}} D(\mathbf{\hat{x}})}_2-1 \right)^2\right].}_\text{Gradient penalty}
    \label{eq:wgan-gp}
\end{equation}
Here, the samples $\mathbf{\hat{x}}$ from the distribution $\mathbb{P}_\mathbf{\hat{x}}$ are obtained by interpolating uniformly along straight lines between pairs of points sampled from the data distribution $\mathbb{P}_r$ and the generator distribution $\mathbb{P}_g$.

In our first approach, we train a WGAN with gradient penalty (WGAN-GP) according to Eq.~\ref{eq:wgan-gp} as proposed in \cite{gulrajani2017improved}. 
The details of training and the obtained results are in Section~\ref{sec:exp}.

\begin{figure}[t]
\begin{center}
\setlength{\tabcolsep}{5pt}
\begin{tabular}{cc}
  \includegraphics[width=0.45\linewidth]{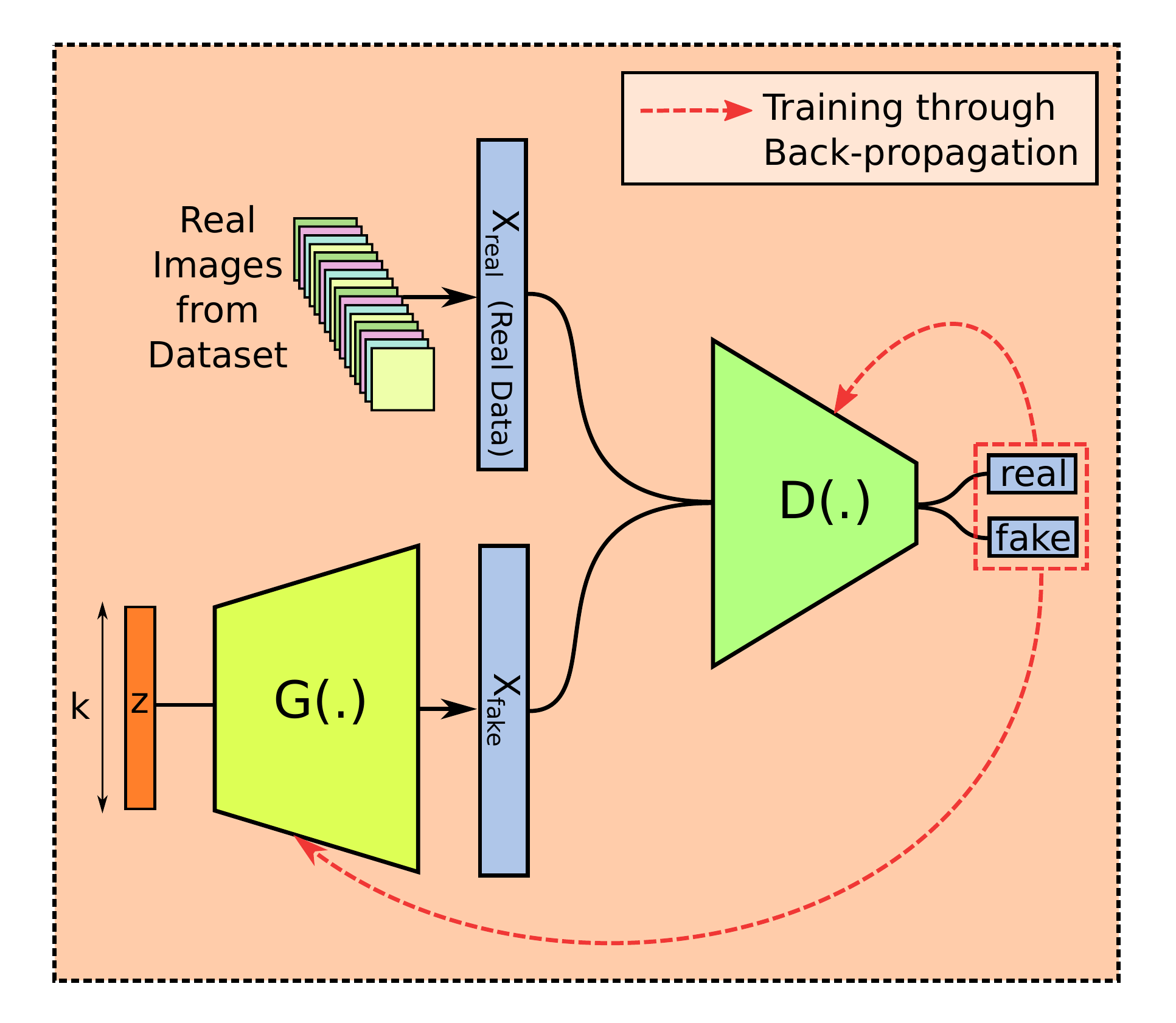}
     & 
  \includegraphics[width=0.45\linewidth]{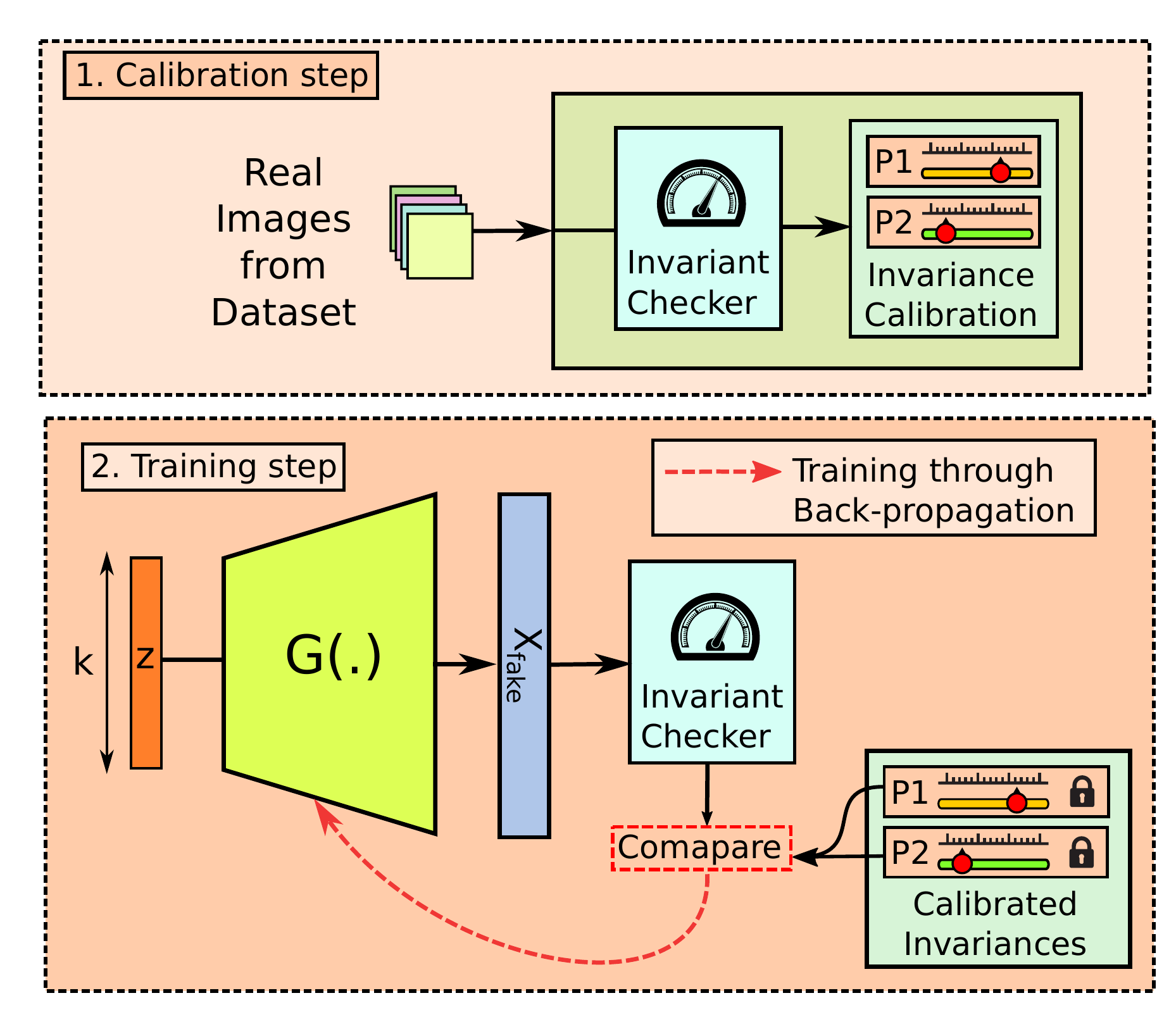} \\
    (a)  & (b)
\end{tabular}
\caption{(a) Generative Adversarial Network model; (b) Generative Invariance Network model}
\label{fig:dias}
\end{center}
\end{figure}

\subsection{Generative Invariance Network}
\label{methods:invarGAN}

Though our WGAN-GP model shows an ability to learn the statistical descriptors of the given microstructure morphologies to some extent, it gives very coarse control on the properties of the generated samples. The only factor determining the generated samples is the training data, making it difficult for the model to generate samples with precise values of $p_1$, $p_2$ and other higher order statistical measures on demand. Such a need usually arises in multiple fields of material science. 
To resolve this, we advocate a physics-based generative model that we call the {\bf Generative Invariance Network} (GIN).


Our proposed GIN model consists of two parts, illustrated in Fig.~\ref{fig:dias}(b). The key innovation is to replace the discriminator in a traditional GAN model with a function that verifies whether the generated samples obey known physical invariances. We use such a mathematical description the \emph{Invariance Checker} function. 

The first step in training our model is the \emph{calibration step}. Here, the invariance checker estimates, from the training data, the parameters of the invariances that need to be verified. For example, in our application, the checker calculates the target $p_1$ and $p_2$ values for all the training image, and fine-tunes (calibrates) threshold parameters $p^{*}_1$ and $p^{*}_2$.
The second step is standard back-propagation, where the calibrated invariance checker produces a loss value that can be used to train the weights of the generator. More specifically, the checker calculates the $p_1$ and $p_2$ values for each image generated by the generator (G), and compares it against the thresholds set in the calibration step. A mismatch in these values can be used to define the loss. In the simplest form, this can be some sort of weighted $\ell_2$-loss, defined as:
\begin{equation}
L_{inv} = \lambda_1\sum_i |p^{(g)}_1 - p^{*}_1 | + \lambda_2\sum_i \|p^{(g)}_2 - p^{*}_2 \|_2,~~g \in \text{set of images generated by generator,}~\mathcal{G}.
\label{methods:invar-loss-eq}    
\end{equation}
This loss can be back-propagated to train the generator weights. After sufficient training, generator is expected to produce images with statistical properties closely matching with the calibrated invariances. 
In Eq.~\ref{methods:invar-loss-eq}, the parameters $\lambda_1$ and $\lambda_2$ can be used to trade-off the fidelity of the generated samples to the target invariances. Therefore, we obtain fine-grained control over the physical properties fo the outputs, compared to traditional GANs.

The benefits of being able to incorporate physical invariances into generative model training are two-fold. First, we obtain superior control over features of the target images. Second, there is no requirement of training a full-fledged discriminator, but rather, only calibrating the parameters of the invariance-checker. That means that such generative models can be trained even with substantially less training data.

\subsection{Hybrid (GAN+GIN) model}
We propose a third approach that combines the best aspects of both the previous models into a Hybrid (GAN+GIN) model. 

Similar to GIN, the hybrid model too proceeds in two steps: a calibration step followed by training step. Calibration step is essentially the same as the GIN, where the invariances are calibrated. However, in the training step, instead of solely relying on either discriminator or the invariance checker, we use a combination of both to train the generator. Two separate optimizers are run to train the generator, one each for minimizing the discriminator loss and the invariant loss. The optimization procedures remain same as in the cases of GAN and GIN respectively. Similar to GAN, we update the generator and discriminator alternatively, with the generator getting updated two times in every iteration: once for minimizing discriminator loss, and once for minimizing the invariance loss~\eqref{methods:invar-loss-eq}. 


\section{Experiments}
\label{sec:exp}
\subsection{Cahn-Hilliard Dataset}
We use the Cahn-Hilliard equation~\cite{cahn1958free} to generate microstructures for training and testing. Originally proposed to study phase separation in alloys, this equation can account for such phenomena in polymers, ceramics, and other material systems. We use an in-house modular finite element software~\cite{Wodo2012} to solve this equation to generate time evolving microstructures. 
Independent solutions were obtained for a range of system parameters like volume fraction (ratio of black pixels to white pixels) and immiscibility parameters (that determine the degree of purity of domains). The simulation domains are square in shape and have physical dimensions of 100 units each. This data set is a modified version of the data used for~\cite{Wodo2015}.

\subsection{Experiments on WGAN-GP model}

\subsubsection{Architecture}\label{sec:wgan-arch} 
In all of our WGAN-GP experiments, we use the ResNet-type architecture from \cite{gulrajani2017improved} with some modifications. In our architecture, both the generator and the discriminator are made of $4$ residual blocks, each block consisting two $3 \times 3$ convolutional layers with the ReLU nonlinearity. Input to generator is $128-$dimensional random Gaussian vector $\mathbf{z}$. In the generator, some residual blocks perform nearest neighbor upsampling, while in the case of discriminator, some residual blocks perform downsampling through average pooling. Batch normalization is applied only in the case of generator. For optimization, we use Adam 
with learning rate $1 \times 10^{-4}$, and batch size $32$. Dimensions of the input images and output samples are $64 \times 64 \times 1$. The hyper-parameter for gradient penalty term, $\lambda$ in Eq.~\ref{eq:wgan-gp} is set to $10$. The number of discriminator updates per generator update is set to $5$.
\begin{figure}[t]
\begin{center}
\setlength{\tabcolsep}{3pt}
 \renewcommand{\arraystretch}{0.9}
\begin{tabular}{cc}
  \includegraphics[width=0.45\linewidth]{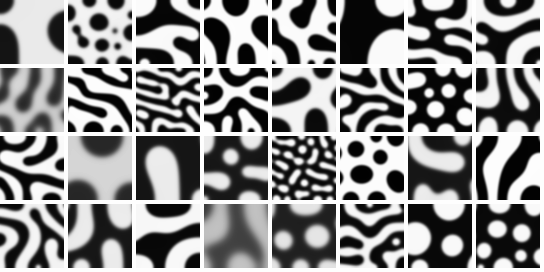}
     & 
  \includegraphics[width=0.45\linewidth]{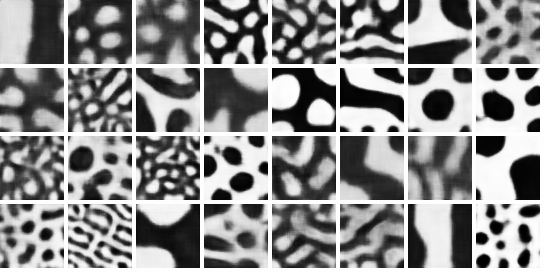} \\
    (a)  & (b) \\
     \includegraphics[width=0.45\linewidth]{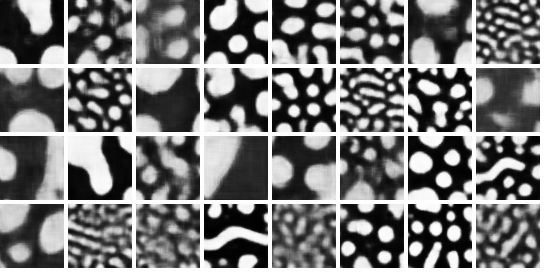}
     & 
  \includegraphics[width=0.45\linewidth]{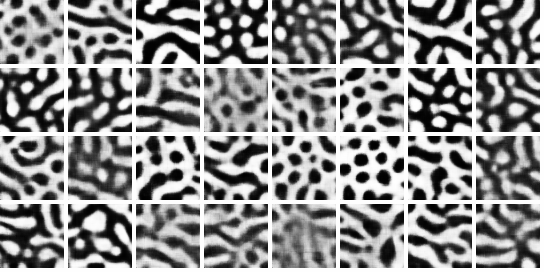} \\
    (c)  & (d) \\
\end{tabular}
\caption{(a) Sample images from Cahn-Hilliard dataset; (b) samples generated by WGAN-GP trained on CH-dataset; (c) Samples generated by WGAN-GP trained over the morphologies from $\mathrm{CH}_{p1}$ dataset (only includes the images with volume fraction between $0.35$ to $0.45$); (d) Samples generated by WGAN-GP trained over the morphologies from $\mathrm{CH}_{p2}$ dataset (only includes the images with $2-$point correlation equal to $0.0625$).}
\label{fig:wgan-samples}
\end{center}
\end{figure}

\subsubsection{Qualitative results and comparisons} 

Using the Cahn-Hilliard (CH) dataset, we prepare two smaller datasets referred as $\mathrm{CH}_{p1}$ and $\mathrm{CH}_{p2}$ by segregating the images based on their $p_1$ and $p_2$ values respectively. The first subset $\mathrm{CH}_{p1}$ is a collection of all images of CH dataset having a volume fraction ($p_1$) value between $0.35$ to $0.45$. The second subset is segregated on the basis of $p_2$ values of images, and contain all the images from CH dataset having $2-$point correlation ($p_2$) value equal to $0.0625$. We train $3$ WGAN-GP using $\mathrm{CH}$, $\mathrm{CH}_{p1}$ and $\mathrm{CH}_{p2}$ as the training data respectively. As these segregated datasets typically contain images with similar statistical properties (either $p_1$ or $p_2$), we can testify the ability of our model to preserve such properties by observing the $p_1$ or $p_2$ values of the images generated by these $3$ networks.

\begin{figure}[t]
\begin{center}
\setlength{\tabcolsep}{2pt}
\begin{tabular}{ccc}
  \begin{tikzpicture}[scale=0.535]
  \tikzstyle{every node}=[font=\Large]
  \begin{axis}[
  ymin=0,axis background/.style={fill=seabornback, fill opacity=1},
  grid style={line width=.1pt, draw=white},
  major grid style={line width=.1pt,draw=white},
  minor tick num=1,
  xmin=0.2,
  xmax=0.8,
  xlabel=Volume fraction,
  ylabel=Sample density,
  grid=both,
  legend cell align=left,
  legend style={at={(1,1.25)}},ylabel style={at={(2ex,0.5)}}
  ]
\addplot[thick,color=ube,fill, fill opacity=0.2,area legend] table[x=x, y=kde, col sep=comma]{data/global_vol_frac_kde.csv};
\addplot[thick,color=amber,fill, fill opacity=0.2,area legend] table[x=x, y=kde, col sep=comma]{data/global_gen_vol_frac_kde.csv};
\legend{Real data distribution, Generated data distribution}
\end{axis}
\end{tikzpicture}
     & 
     \begin{tikzpicture}[scale=0.535]
     \tikzstyle{every node}=[font=\Large]
  \begin{axis}[ymin=0,axis background/.style={fill=seabornback, fill opacity =1},
grid style={line width=.1pt, draw=white},
    major grid style={line width=.1pt,draw=white},
    minor tick num=1,
     grid=both,
     xmin=0.2,
  xmax=0.8,
     xlabel=Volume fraction,
     ylabel=Sample density,
     legend cell align=left,
     legend style={at={(1,1.25)}},ylabel style={at={(2ex,0.5)}}]
\addplot[thick,color=ube,fill, fill opacity=0.2,area legend] table[x=x, y=kde, col sep=comma]{data/sec_p1_kde_real.csv};
\addplot[thick,color=amber, fill, fill opacity=0.2,area legend] table[x=x, y=kde, col sep=comma]{data/sec_p1_kde_gen.csv};
\legend{Real data distribution, Generated data distribution}
\end{axis}
\end{tikzpicture} & 
\begin{tikzpicture}[scale=0.535]
\tikzstyle{every node}=[font=\Large]
 \begin{axis}[
        xmode=linear,
        ymode=linear,
        ymin=0,axis background/.style={fill=seabornback, fill opacity =1},
    grid style={line width=.1pt, draw=white},
    major grid style={line width=.1pt,draw=white},
    minor tick num=1,
     grid=both,
     legend cell align=left,
     legend style={at={(1,1.25)}},
     xlabel=$p_2$ value,
     ylabel=Number of samples,
      xticklabel style={
        /pgf/number format/fixed,
        /pgf/number format/precision=4},
     scaled ticks=base 10:-3,
        scaled x ticks=false,
        ylabel style={at={(1ex,0.5)}}]
    
    \addplot+[thick, color=ube, no marks, area legend, ybar, bar width=10, fill, fill opacity=0.4] plot coordinates
	{(0.0625,1024)};
\addplot+[thick, color=amber, no marks, area legend, ybar, bar width=10, fill, fill opacity=0.4] plot coordinates
	{(0.046875,52) (0.0625,961) (0.078125,11)};
    
     \legend{Real data, Generated data} 
\end{axis}
 
\end{tikzpicture} \\
    (a)  & (b) & (c)
\end{tabular}
\caption{(a) Comparisons of the distributions of volume fractions of training dataset and that of the samples generated by WGAN-GP trained over entire CH dataset; (b) Comparisons of the distributions of volume fractions of training dataset and that of the samples generated by WGAN-GP trained over the $\mathrm{CH}_{p1}$ dataset; (c) Histograms of $p_2$ correlation values for samples from $\mathrm{CH}_{p2}$ dataset and samples generated by WGAN-GP trained over $CH_{p2}$ dataset.} 
\label{fig:hist_p1_p2}
\end{center}
\end{figure}
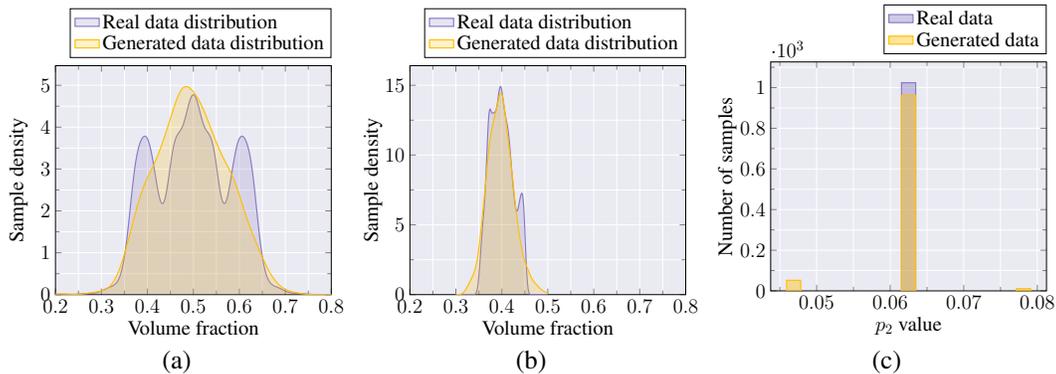

To validate our model's capability of correctly translating the statistical properties of the training data into the generated images, we provide various qualitative results for $3$ WGAN-GP networks each trained on $\mathrm{CH}$, $\mathrm{CH}_{p1}$ and $\mathrm{CH}_{p2}$ respectively. The model architecture and other hyperparameters are kept same for all $3$ networks. In Fig.~\ref{fig:wgan-samples} (b), (c) and (d) we depict the samples generated by all $3$ networks respectively. Qualitatively, we can conclude that these results maintain the diversity within the dataset, while preserving the uniformity in the statistical properties. We further provide the density plots/histograms of $p_1$ or $p_2$ values of the images for both real data and generated data. The striking similarities in the spread of both the density plots/histograms suggest that our network successfully reproduces the statistical properties of the real (training) images in the simulated images. In Fig.~\ref{fig:hist_p1_p2} (a,b), densities of $p_1$ value is compared between the real data and generated data for network trained on $\mathrm{CH}$ dataset and $\mathrm{CH}_{p1}$ dataset. 
We provide the histogram for $p_2$ values for models trained using $\mathrm{CH}_{p2}$ in Fig.~\ref{fig:hist_p1_p2} (c). Both the histograms closely match.  
\subsubsection{Interpolation results}
\begin{figure}[t]
	\begin{center}
		\setlength{\tabcolsep}{1pt}
		\begin{tabular}{cccccccccccc}
		\includegraphics[width=\w\linewidth, trim={\sa em \s em 0 0},clip]{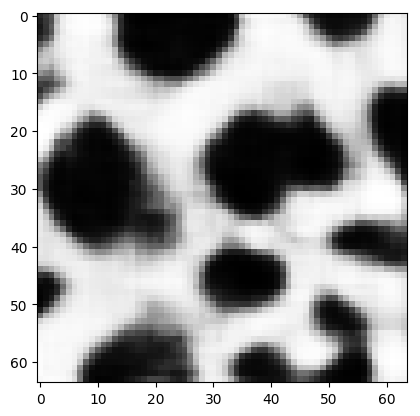} &
		\includegraphics[width=\w\linewidth, trim={\sa em \s em 0 0},clip]{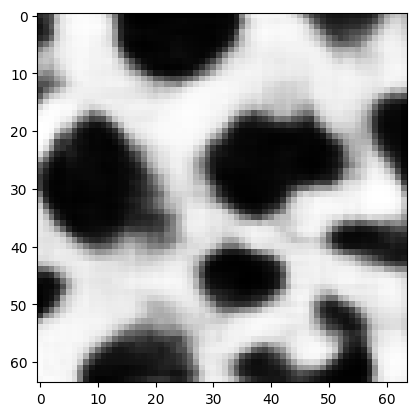} &
		\includegraphics[width=\w\linewidth, trim={\sa em \s em 0 0},clip]{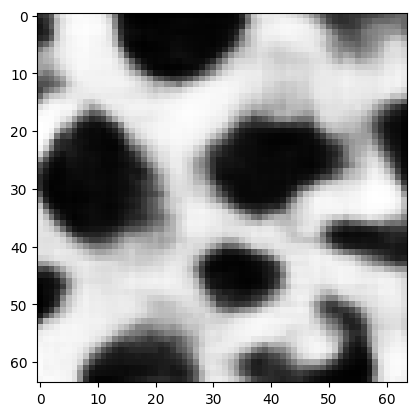} &
		\includegraphics[width=\w\linewidth, trim={\sa em \s em 0 0},clip]{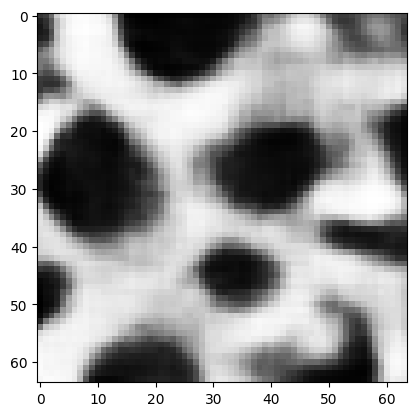} &
		\includegraphics[width=\w\linewidth, trim={\sa em \s em 0 0},clip]{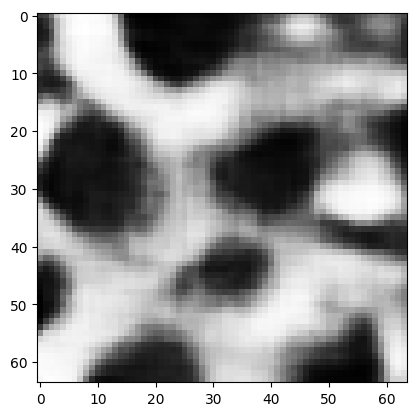} &
		\includegraphics[width=\w\linewidth, trim={\sa em \s em 0 0},clip]{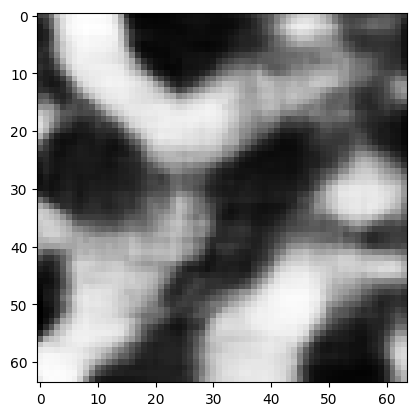} &
		\includegraphics[width=\w\linewidth, trim={\sa em \s em 0 0},clip]{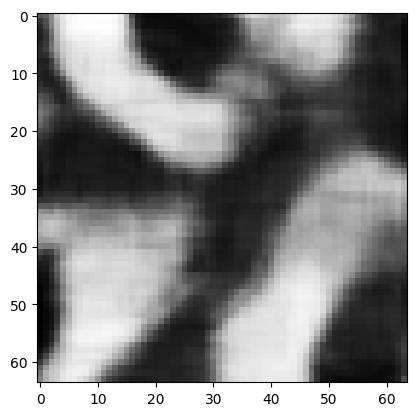} &
		\includegraphics[width=\w\linewidth, trim={\sa em \s em 0 0},clip]{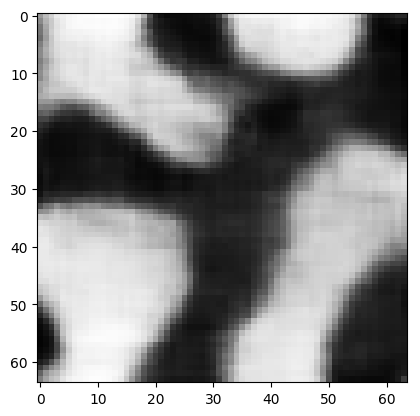} &
		\includegraphics[width=\w\linewidth, trim={\sa em \s em 0 0},clip]{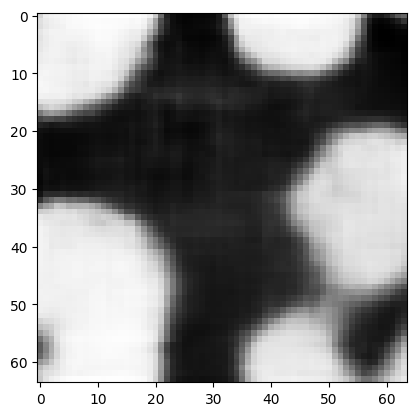} &
		\includegraphics[width=\w\linewidth, trim={\sa em \s em 0 0},clip]{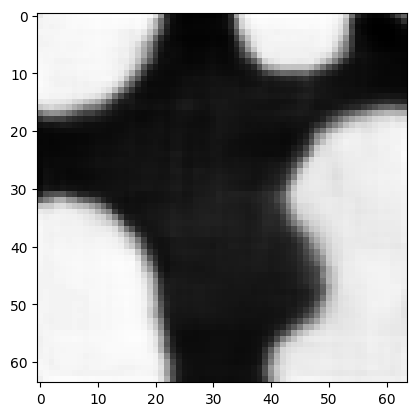} &
		\includegraphics[width=\w\linewidth, trim={\sa em \s em 0 0},clip]{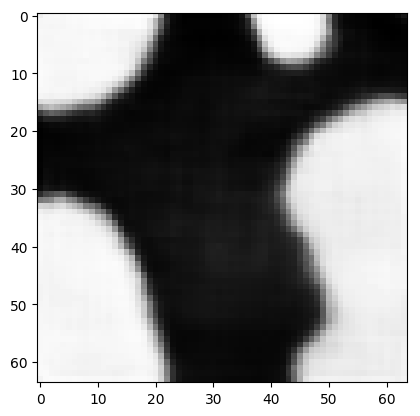}&
		\includegraphics[width=\w\linewidth, trim={\sa em \s em 0 0},clip]{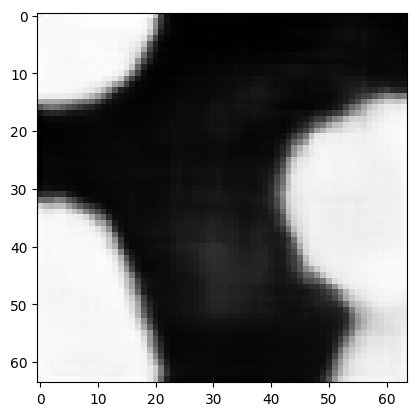} \\
		
		\includegraphics[width=\w\linewidth,trim={\sa em \s em 0 0},clip]{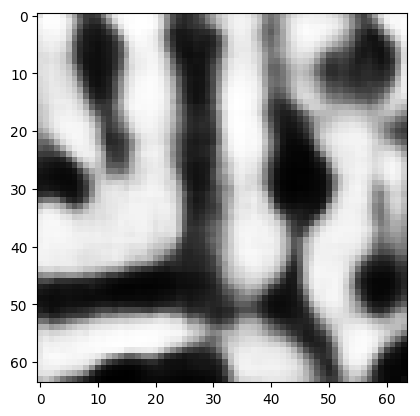} &
		\includegraphics[width=\w\linewidth,trim={\sa em \s em 0 0},clip]{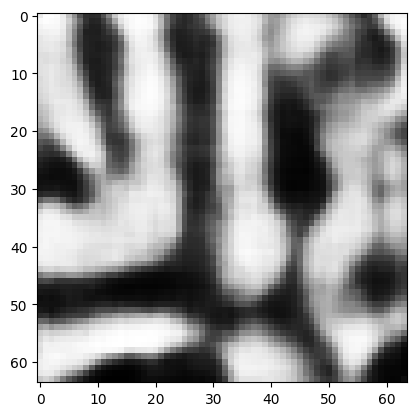} &
		\includegraphics[width=\w\linewidth,trim={\sa em \s em 0 0},clip]{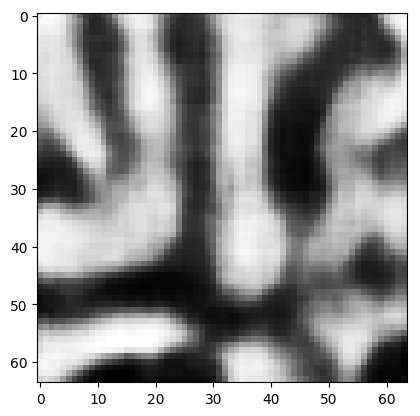} &
		\includegraphics[width=\w\linewidth,trim={\sa em \s em 0 0},clip]{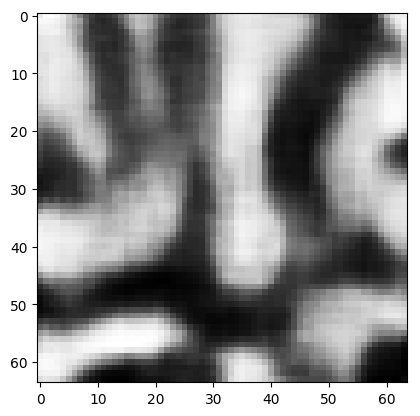} &
		\includegraphics[width=\w\linewidth,trim={\sa em \s em 0 0},clip]{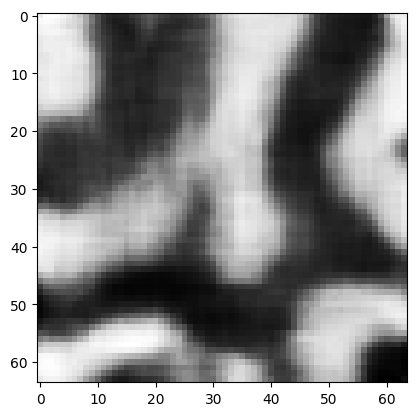} &
		\includegraphics[width=\w\linewidth,trim={\sa em \s em 0 0},clip]{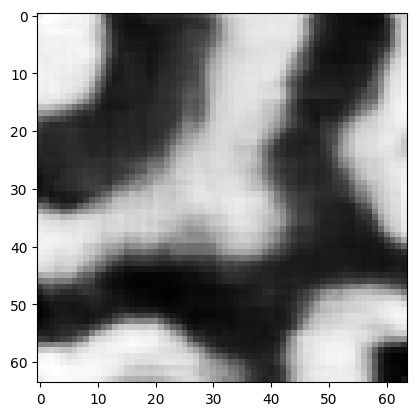} &
		\includegraphics[width=\w\linewidth,trim={\sa em \s em 0 0},clip]{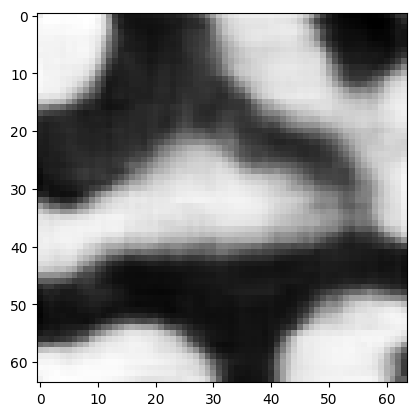} &
		\includegraphics[width=\w\linewidth,trim={\sa em \s em 0 0},clip]{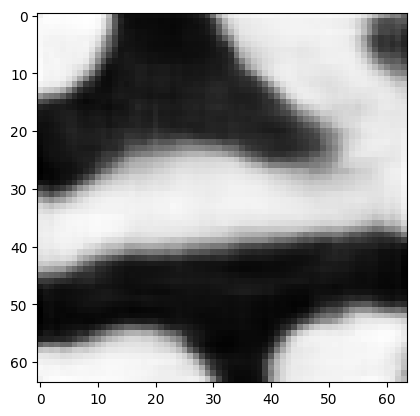} &
		\includegraphics[width=\w\linewidth,trim={\sa em \s em 0 0},clip]{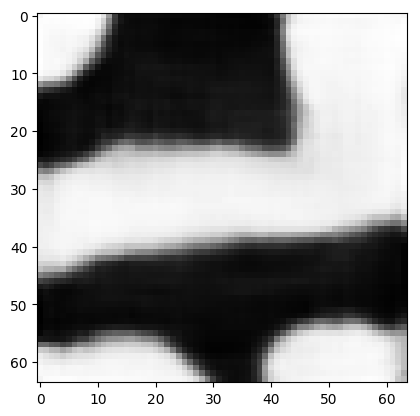} &
		\includegraphics[width=\w\linewidth,trim={\sa em \s em 0 0},clip]{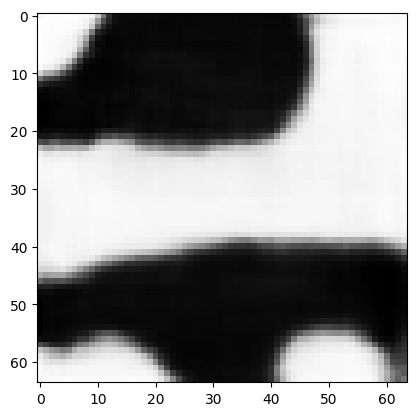} &
		\includegraphics[width=\w\linewidth,trim={\sa em \s em 0 0},clip]{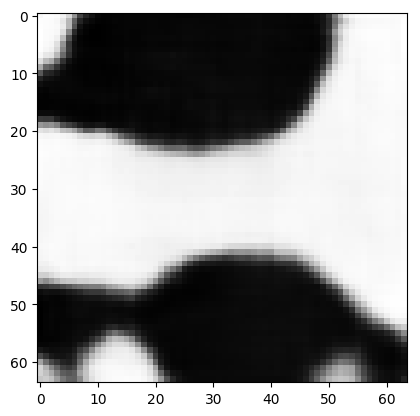} &
		\includegraphics[width=\w\linewidth,trim={\sa em \s em 0 0},clip]{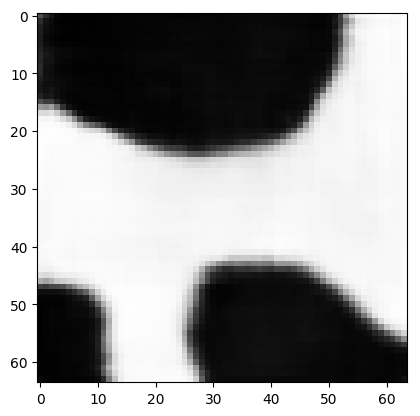} \\ 
	\end{tabular}
	\end{center}
	\caption{{(a) Results of the linear interpolation over latent variable $z$ for WGAN-GP trained over the entire CH dataset}}
	\label{fig:global_interp}
\end{figure}

Further, we display interesting behavior of the learned image manifold through results of interpolation over latent vectors $\mathbf{z}$. For the first WGAN-GP network trained over the full CH dataset, we randomly pick two different noise vectors $\mathbf{z}_1$ and $\mathbf{z}_2$, and linearly interpolate between them to obtain $10$ more such noise vectors. All $12$ images are plotted as in Fig.~\ref{fig:global_interp}. It is noticeable that the image manifold appears interpretable, providing a smooth transition from one morphology to another.

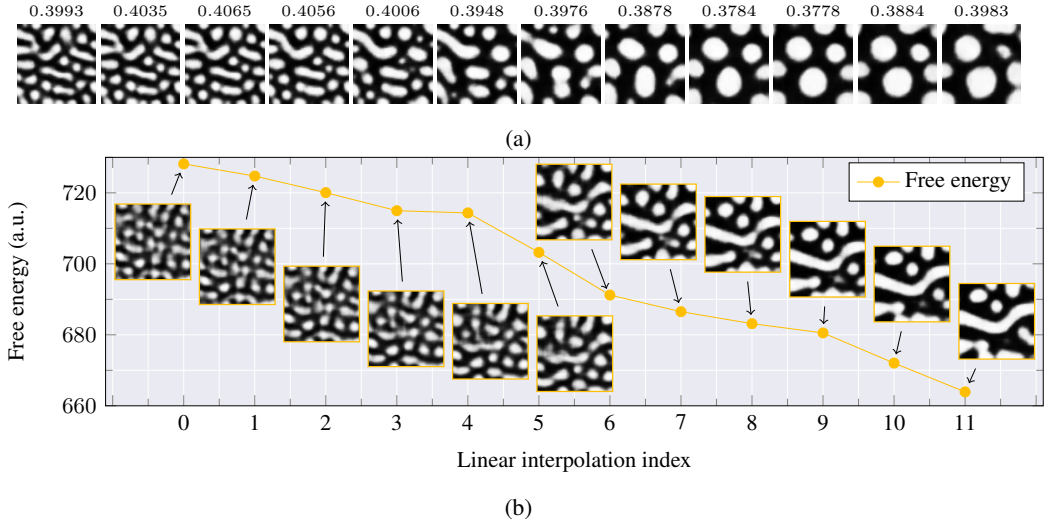
\begin{figure}[t!]
	\begin{subfigure}[t]{\linewidth}
		\centering
		\setlength{\tabcolsep}{1pt}
		\begin{tabular}{cccccccccccc}
		\tiny{$0.3993$} &
		\tiny{$0.4035$} &
		\tiny{$0.4065$} &
		\tiny{$0.4056$} &
		\tiny{$0.4006$} &
		\tiny{$0.3948$} &
		\tiny{$0.3976$} &
		\tiny{$0.3878$} &
		\tiny{$0.3784$} &
		\tiny{$0.3778$} & 
		\tiny{$0.3884$} &
		\tiny{$0.3983$} \\
		
		\includegraphics[width=\wa\linewidth ]{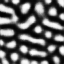} &
		\includegraphics[width=\wa\linewidth ]{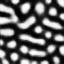} &
		\includegraphics[width=\wa\linewidth ]{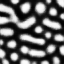} &
		\includegraphics[width=\wa\linewidth ]{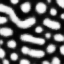} &
		\includegraphics[width=\wa\linewidth ]{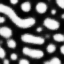} &
		\includegraphics[width=\wa\linewidth ]{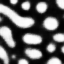} &
		\includegraphics[width=\wa\linewidth ]{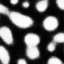} &
		\includegraphics[width=\wa\linewidth ]{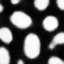} &
		\includegraphics[width=\wa\linewidth ]{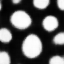} &
		\includegraphics[width=\wa\linewidth ]{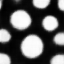} &
		\includegraphics[width=\wa\linewidth ]{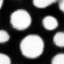} &
		\includegraphics[width=\wa\linewidth ]{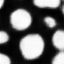} \\ 
    
	\end{tabular}
	\caption{}
	\end{subfigure}
	\begin{subfigure}[b]{\linewidth}
	          \centering
	  \tikzstyle{every node}=[font=\small]
	  \resizebox{\linewidth}{!}{
    \begin{tikzpicture}
        \begin{axis}[
        axis background/.style={fill=seabornback, fill opacity=1}, 
        ymin=660, ymax=730,
        grid style={line width=.1pt, draw=white},
        major grid style={line width=.1pt,draw=white},
        minor tick num=1,
        grid=both,
        x=1cm,
        y=0.05cm,
        xticklabels={, ,0,1,2,3,4,5,6,7,8,9,10,11},
        xlabel=Linear interpolation index,
        ylabel= Free energy (a.u.)
        ]
        \addplot+[color=amber,mark=*, mark options={fill=amber, solid}]
            coordinates{
            (0,728.158008855962)
            (1,724.730815230967)
            (2,720.060098555796)
            (3,714.971928127032)
            (4,714.353220591809)
            (5,703.228007692119)
            (6,691.200090244845)
            (7,686.518737585112)
            (8,683.151706704045)
            (9,680.535518510516)
            (10,672.038832742987)
            (11,663.917988503618)
            };
            \node (source1)[below right] at (rel axis cs:0.0,0.85){\includegraphics[width=\wa\linewidth, cframe=amber 0.5pt]{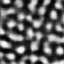}};
            \node (destination1) at (axis cs:0,728.158008855962){};
            \node (source2)[below right] at (rel axis cs:0.09,0.75){\includegraphics[width=\wa\linewidth, cframe=amber 0.5pt]{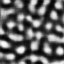}};
            \node (destination2) at (axis cs:1,724.730815230967){};
             \node (source3)[below right] at (rel axis cs:0.18,0.60){\includegraphics[width=\wa\linewidth, cframe=amber 0.5pt]{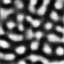}};
             \node (destination3) at (axis cs:2,720.060098555796){};
            \node (source4)[below right] at (rel axis cs:0.27,0.50){\includegraphics[width=\wa\linewidth, cframe=amber 0.5pt]{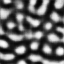}};
            \node (destination4) at (axis cs:3,714.971928127032){};
            \node (source5)[below right] at (rel axis cs:0.36,0.45){\includegraphics[width=\wa\linewidth, cframe=amber 0.5pt]{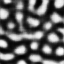}};
            \node (destination5) at (axis cs:4,714.353220591809){};
            \node (source6)[below right] at (rel axis cs:0.45,0.40){\includegraphics[width=\wa\linewidth, cframe=amber 0.5pt]{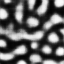}};
             \node (destination6) at (axis cs:5,703.228007692119){};
            \node (source7)[above left] at (rel axis cs:0.55,0.63){\includegraphics[width=\wa\linewidth, cframe=amber 0.5pt]{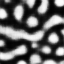}};
            \node (destination7) at (axis cs:6,690.5 00090244845){};
            \node (source8)[above left] at (rel axis cs:0.64,0.55){\includegraphics[width=\wa\linewidth, cframe=amber 0.5pt]{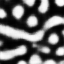}};
            \node (destination8) at (axis cs:7,686.518737585112){};
            \node (source9)[above left] at (rel axis cs:0.73,0.50){\includegraphics[width=\wa\linewidth, cframe=amber 0.5pt]{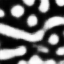}};
            \node (destination9) at (axis cs:8,683.151706704045){};
            \node (source10)[above] at (rel axis cs:0.77,0.40){\includegraphics[width=\wa\linewidth, cframe=amber 0.5pt]{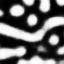}};
            \node (destination10) at (axis cs:9,680.535518510516){};
            \node (source11)[above] at (rel axis cs:0.86,0.30){\includegraphics[width=\wa\linewidth, cframe=amber 0.5pt]{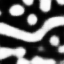}};
            \node (destination11) at (axis cs:10,672.038832742987){};
            \node (source12)[above right] at (rel axis cs:0.90,0.15){\includegraphics[width=\wa\linewidth, cframe=amber 0.5pt]{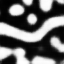}};
            \node (destination12) at (axis cs:11,663.917988503618){};
        \draw[->](source1)--(destination1);
        \draw[->](source2)--(destination2);
        \draw[->](source3)--(destination3);
        \draw[->](source4)--(destination4);
        \draw[->](source5)--(destination5);
        \draw[->](source6)--(destination6);
      \draw[->](source7)--(destination7);
      \draw[->](source8)--(destination8);
      \draw[->](source9)--(destination9);
      \draw[->](source10)--(destination10);
      \draw[->](source11)--(destination11);
      \draw[->](source12)--(destination12);
       \legend{Free energy}
        \end{axis}
        \end{tikzpicture}}
        \caption{}
	\end{subfigure}
	\caption{{(a) Results of the linear interpolation over latent variable $z$ for WGAN-GP trained over the images from $\mathrm{CH}_{p1}$ dataset (only includes the images with volume fraction between $0.35$ to $0.45$); volume fraction values are printed above each image; (b) Free energy of the interpolated morphologies (volume fraction of each image is $(0.41 \pm 5\%)$.}}
	\label{fig:p1_interp}
\end{figure}

For networks trained on $\mathrm{CH}_{p1}$, we repeat similar experiments, except that here we make sure that the initial and final images of the interpolation have nearly the same values of $p_1$ (volume fraction). It is to verify the continuity of the learned manifold with respect to the volume of each material. Fig.~\ref{fig:p1_interp} (a) displays the interpolation results for the $p_1$ case. It is evident that the volume fraction ($p_1$) values remain almost uniform throughout the interpolation. Such behavior can effectively be used to decide series of manufacturing processes for obtaining final morphology from initial morphology without adding any new material. 
Very interestingly, we observe that an unseen invariance of energy minimization is captured. As shown in Fig.~\ref{fig:p1_interp} (b), the free energy of the morphologies decreases as the we move from one interpolation step to the next. In this process, the volume fraction ($p_1$) is also preserved within reasonable limits. This suggests that the WGAN framework is able to learn latent physical rules from the dataset. 

\subsection{Experiments on Generative Invariant Network (GIN)}
\begin{figure}[t]
  \centering
  \includegraphics[width=1.0\linewidth]{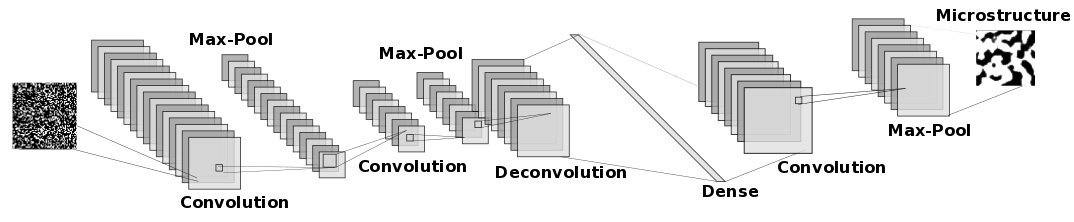}
  \caption{Architecture of the generator (G) for Generative Invariant Network}
  \label{fig:invargan-arch}
\end{figure}
\subsubsection{Architecture}
The input to the generator is a $64\times 64$ square matrix of random noise ($\mathbf{z}$) bounded in $[-1, 1]$. $\mathbf{z}$ is fed into a combination of convolution layer and max-pooling layer. We use two such combinations followed by a batch normalization layer in each combination. The first convolution layer has $32$ filters with a filter size of $4 \times 4$, while the second layer contains $16$ such filters. A max-pooling layer of size $4 \times 4$ with a stride of $2 \times 2$ is applied to each filter, which reduces the size of the representation. It helps in eliminating the dependence in the representations. The output of the two convolutional layers is passed to a deconvolution layer, which has $16$ filters and filter size of $4 \times 4$. Its output is then fed to linear layer of $128 \times 128$ neurons followed by a batch normalization layer. The linear layer collects and aggregates the important information from the local features collected by the previous layers. This is further distributed by means of convolution and max pooling layers with $32$ filters and a filter size of $4 \times 4$. In the final layer, the architecture generates an image of desired size. To maintain consistency between the other models the size of the output images are kept at $64 \times 64$. In the calibration step, a single image from the CH dataset is used for calibrating the GIN invariances. In training step, generator is trained for $15000$ epochs using Adam optimizer with learning rate equal to $10^{-5}$. As $p_2(r = 0) = p_1$, in our training, the loss function in Eq.~\ref{methods:invar-loss-eq} is reduced just to the second term, \textit{i.e.} the term corresponding to $p_2$ value. Further, we consider $p^*_2$ to be a vector containing all $p_2$ values corresponding to the varying distance ($r$). 

\subsubsection{Qualitative results and comparisons}
 Synthetic microstructures are generated using random variables in space bounded in $[-1, 1]$. In Fig.~\ref{fig:invar1} we show the samples generated by three GIN models, each calibrated with a different image from the CH dataset. Real image used for calibration of the invariances are depicted with blue box. Two of these GIN models (a,b) are trained by employing the entire $p_2$ curve of the training image as the invariance, while in the third model (c), we use only the initial part of the $p_2$ curve, as the initial values play a dominating role in deciding the physical and chemical properties of the material.  Visually, generated images have similar domains sizes as the original image and exhibit similar spatial distribution patterns. We plot the $2-$point correlation curves of the generated samples for GIN models and compare it with the $2-$point correlation curve of the real image used for the calibration. Fig.~\ref{fig:invar2a} (a,b) depicts such curves for GIN models (a,b) trained with entire $p_2$ curve. Fig.~\ref{fig:invar2a} (c) depicts such curve for GIN model (c) trained with only the initial values of the $p_2$ curve. It is evident that the $p_2$ curves of the generated images conform to the calibrated $p_2$ curves closely. In es sense we capture three descriptors of the microstructure by capturing $2-$point correlation. The lowest order descriptor is the volume fraction, which is equal to $p_1$. In other words composition of the generated microstructures is very close to the original composition and varies in the range $p_1 \pm 5 \%$. The second descriptor is the surface area, which is equal to $\frac{dp_2 (0)}{dr}$ and graphically represents the slope of the curve near $r = 0$. This values is also close with the value of the original image in all $3$ cases. 

\renewcommand{\wa}{0.055}
\begin{figure}[t!]
	\begin{center}
		\setlength{\tabcolsep}{1pt}
		\renewcommand{\arraystretch}{0.5}
		\begin{tabular}{ccccccccccccccccc}
	    \raisebox{0.8em}{\small{(a)}} &
		\includegraphics[width=0.054\linewidth,cframe=ube 1pt]{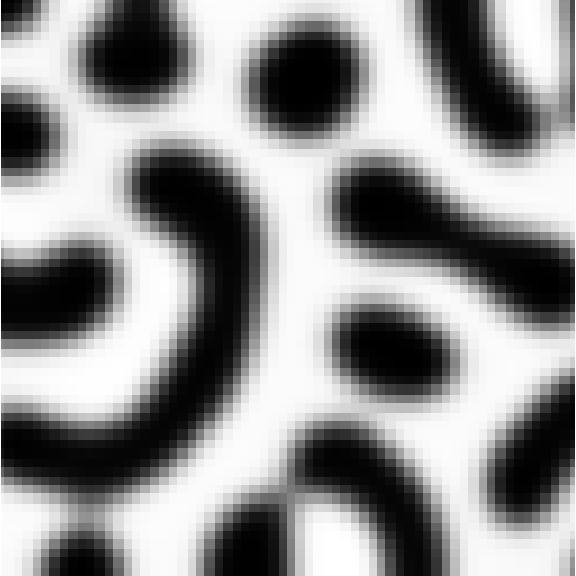} &
		\includegraphics[width=\wa\linewidth]{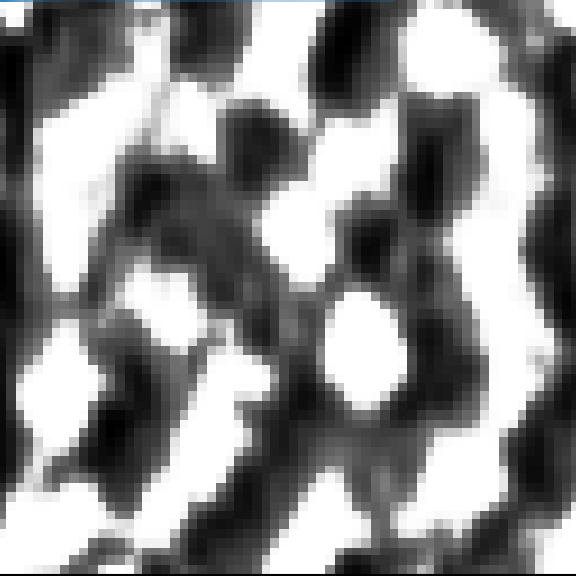} &
		\includegraphics[width=\wa\linewidth]{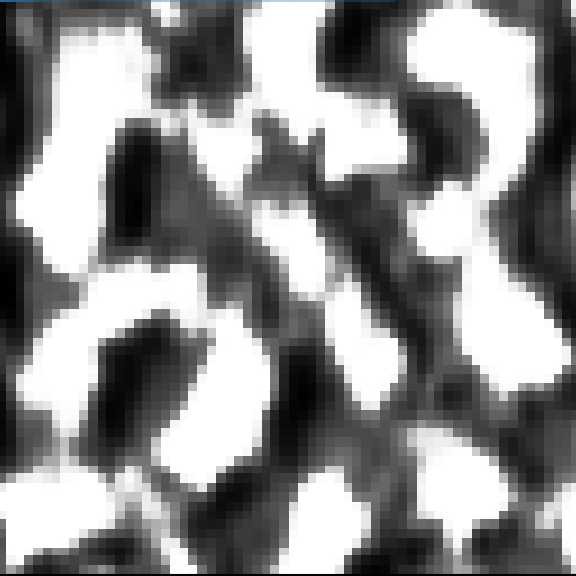} &
		\includegraphics[width=\wa\linewidth ]{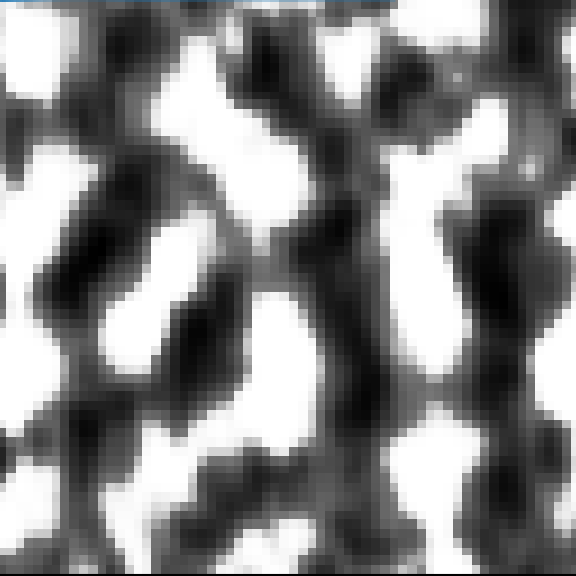} &
		\includegraphics[width=\wa\linewidth ]{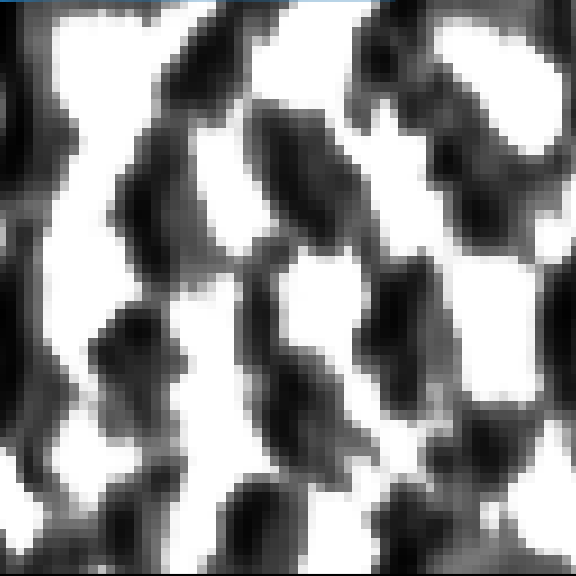} &
		\includegraphics[width=\wa\linewidth ]{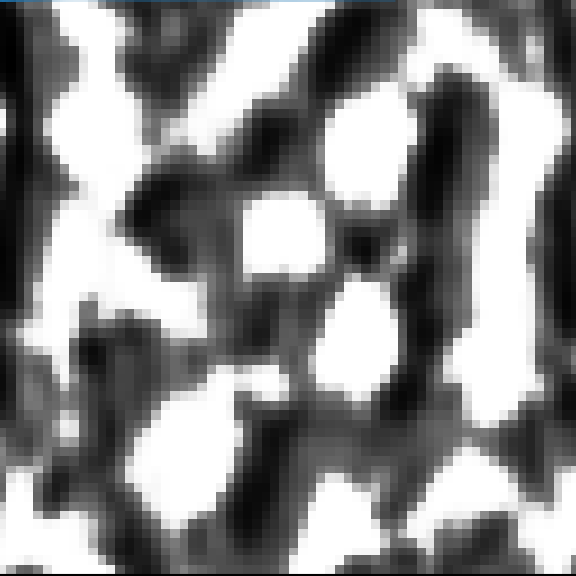} &
		\includegraphics[width=\wa\linewidth ]{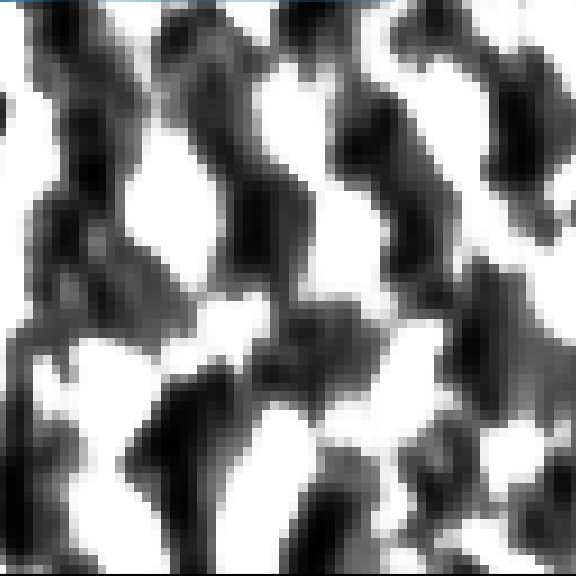} &
		\includegraphics[width=\wa\linewidth ]{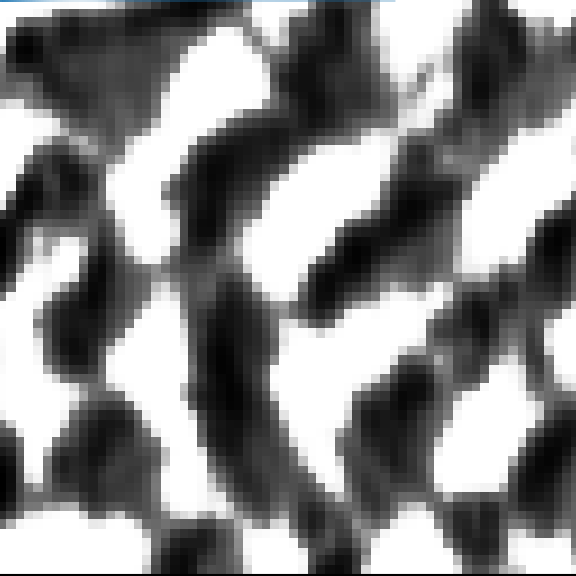} &
		\includegraphics[width=\wa\linewidth ]{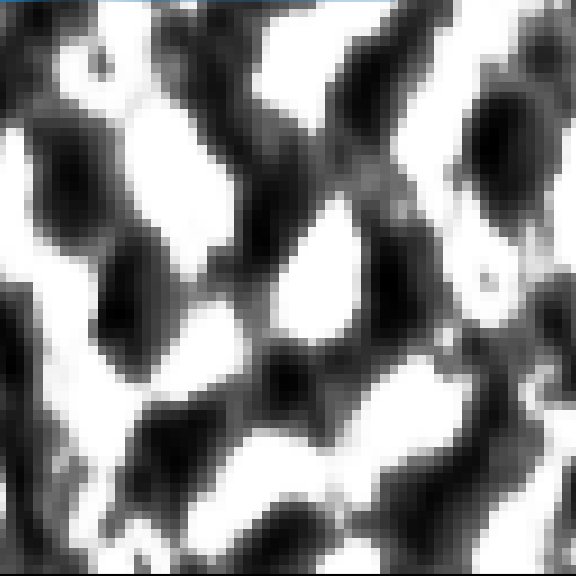} &
		\includegraphics[width=\wa\linewidth ]{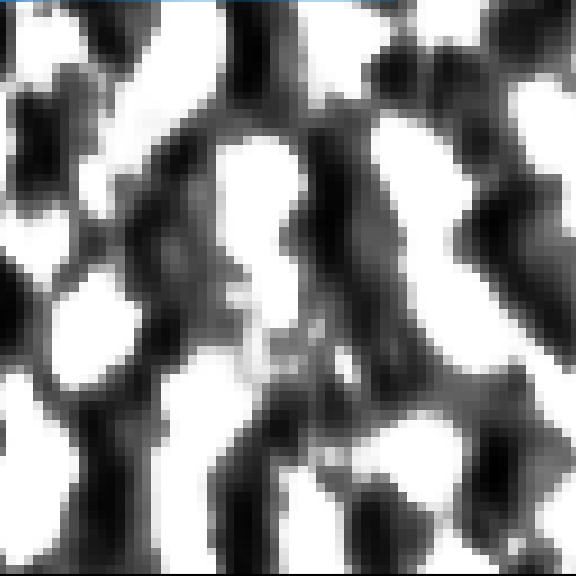} &
		\includegraphics[width=\wa\linewidth ]{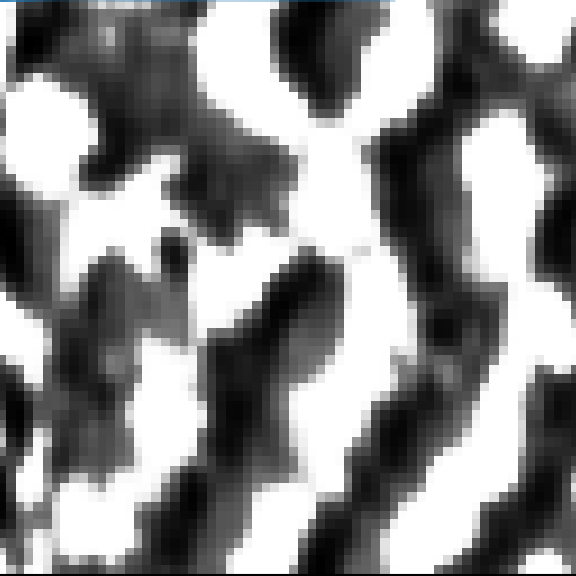} &
		\includegraphics[width=\wa\linewidth ]{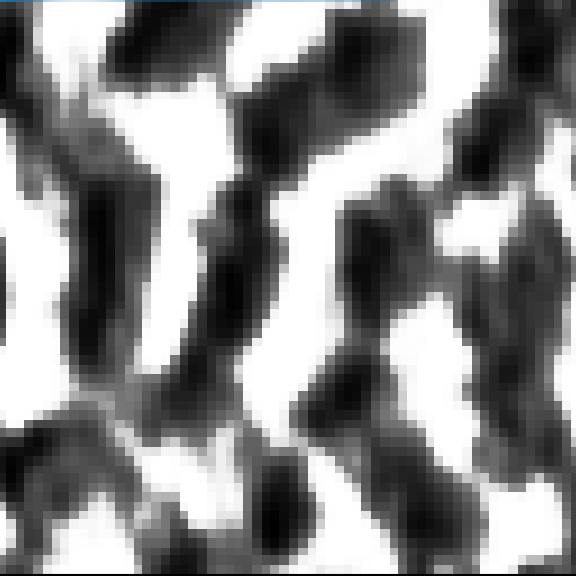} & 
		\includegraphics[width=\wa\linewidth ]{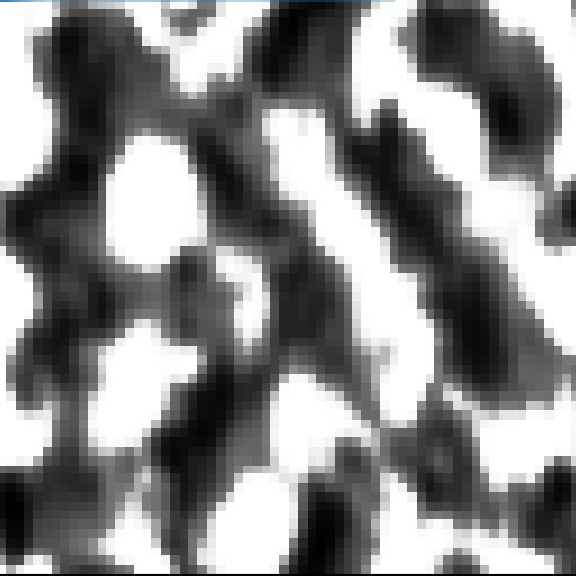} &
		\includegraphics[width=\wa\linewidth ]{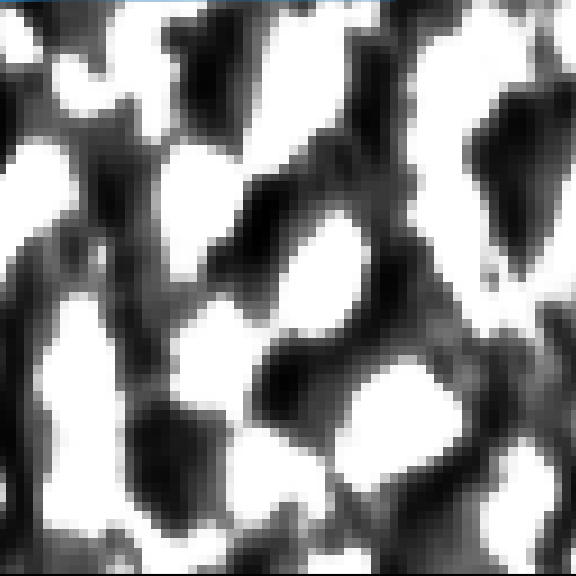} &
		\includegraphics[width=\wa\linewidth ]{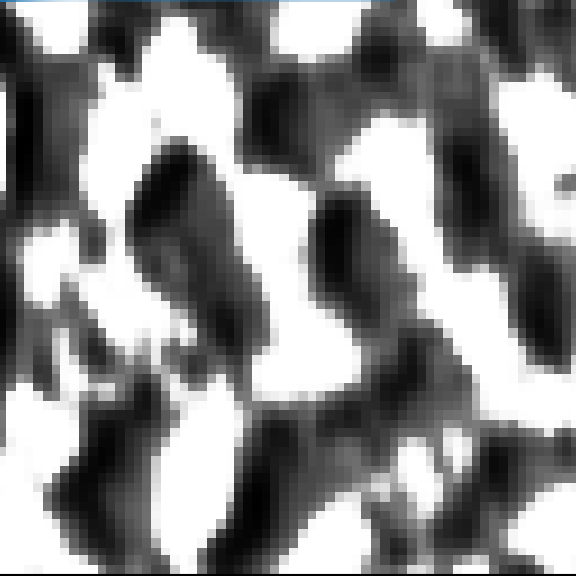} &	
		\includegraphics[width=\wa\linewidth ]{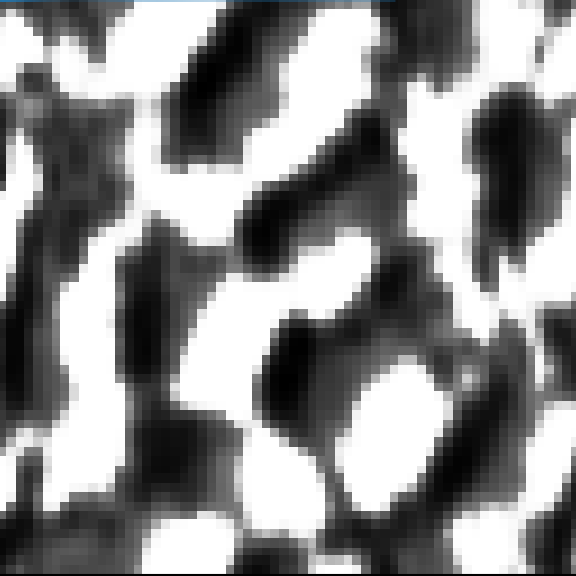} \\ 
	
       \raisebox{0.8em}{\small{(b)}} &
		\includegraphics[width=0.054\linewidth , cframe=ube 1pt]{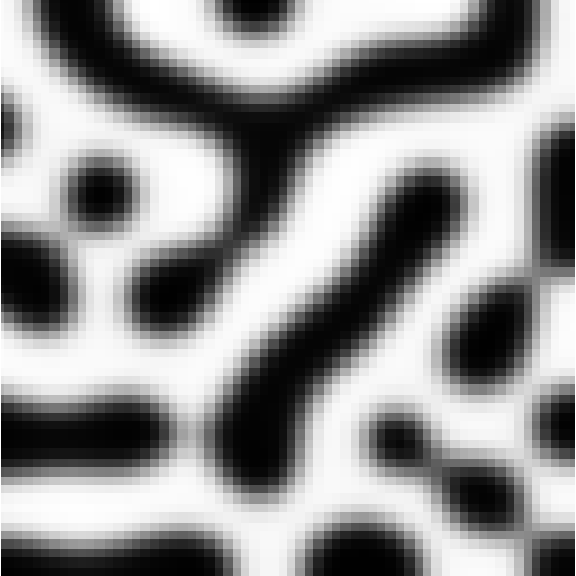} &
		\includegraphics[width=\wa\linewidth ]{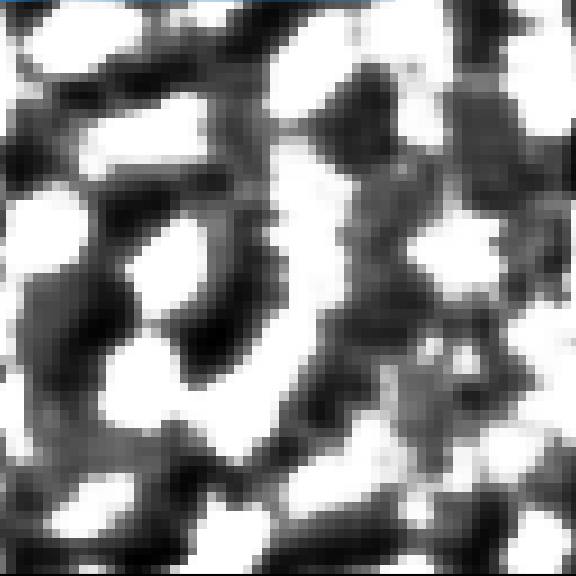} &
		\includegraphics[width=\wa\linewidth ]{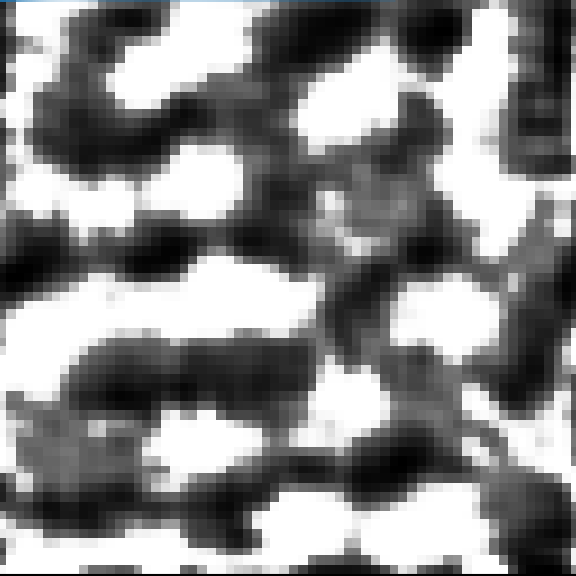} &
		\includegraphics[width=\wa\linewidth ]{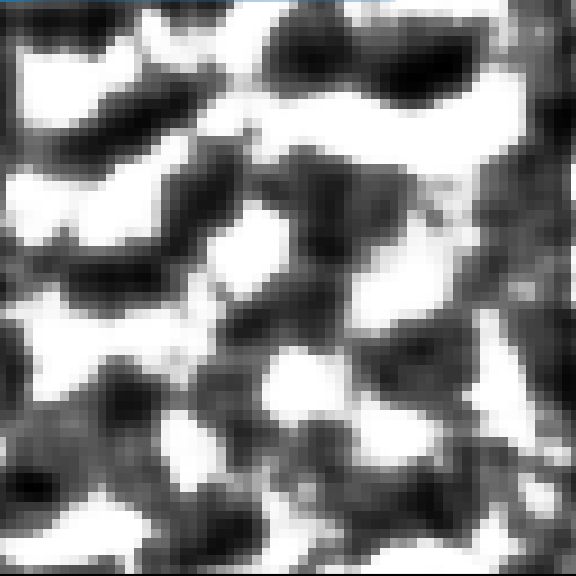} &
		\includegraphics[width=\wa\linewidth ]{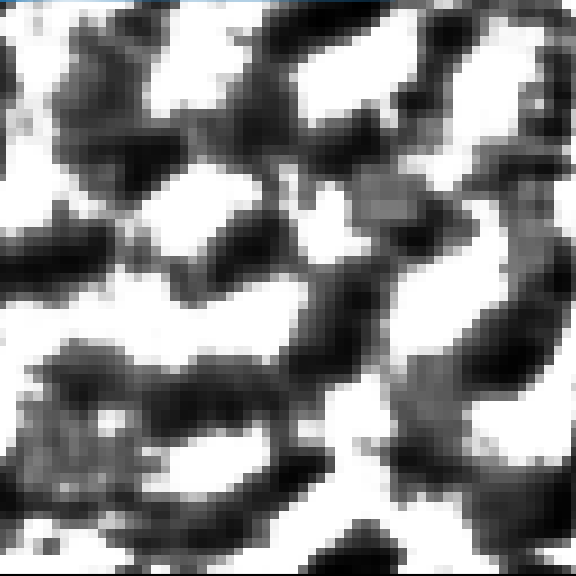} &
		\includegraphics[width=\wa\linewidth ]{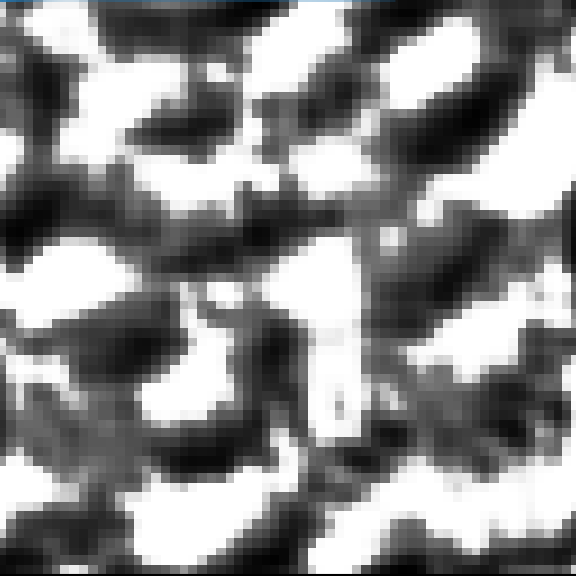} &
		\includegraphics[width=\wa\linewidth ]{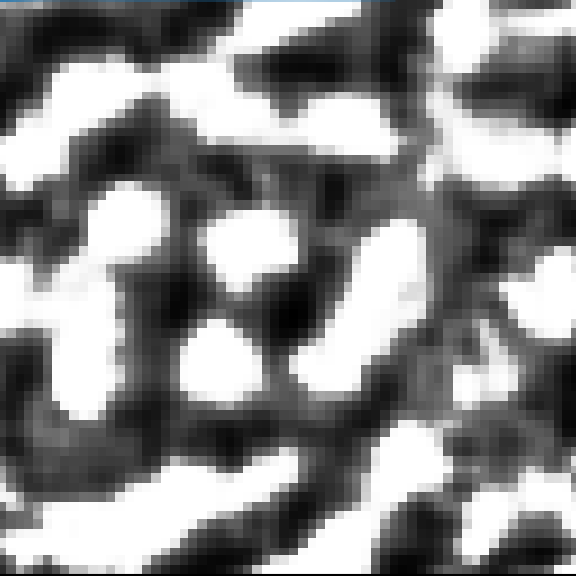} &
		\includegraphics[width=\wa\linewidth ]{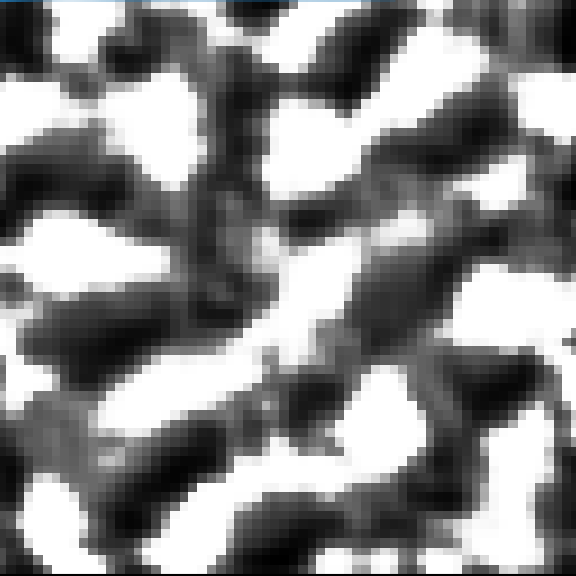} &
		\includegraphics[width=\wa\linewidth ]{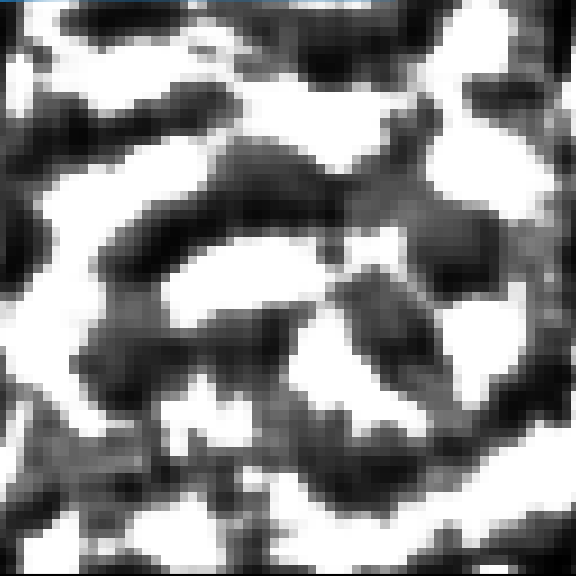} &
		\includegraphics[width=\wa\linewidth ]{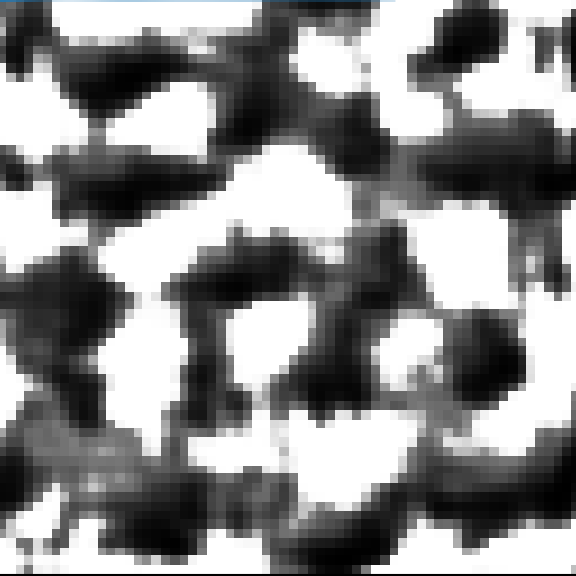} &
		\includegraphics[width=\wa\linewidth ]{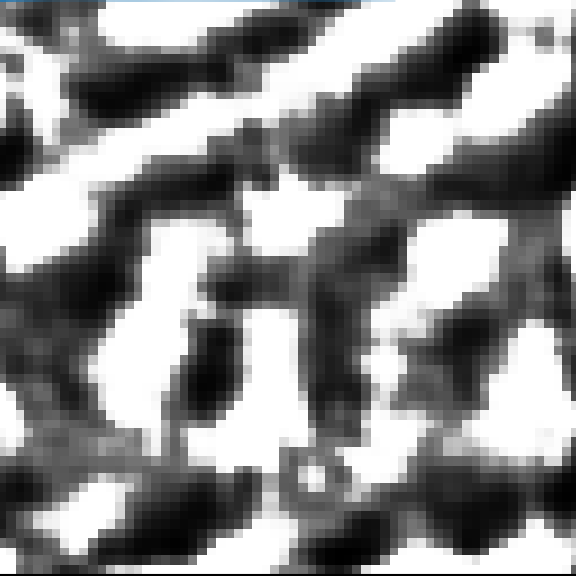} &
		\includegraphics[width=\wa\linewidth ]{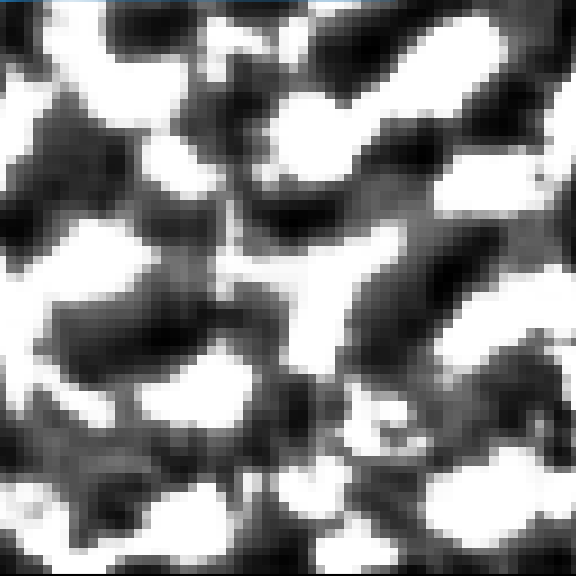} & 
		\includegraphics[width=\wa\linewidth ]{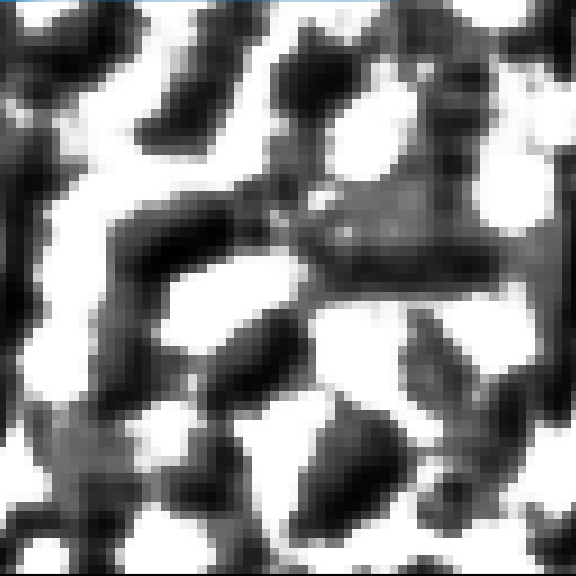} &
		\includegraphics[width=\wa\linewidth ]{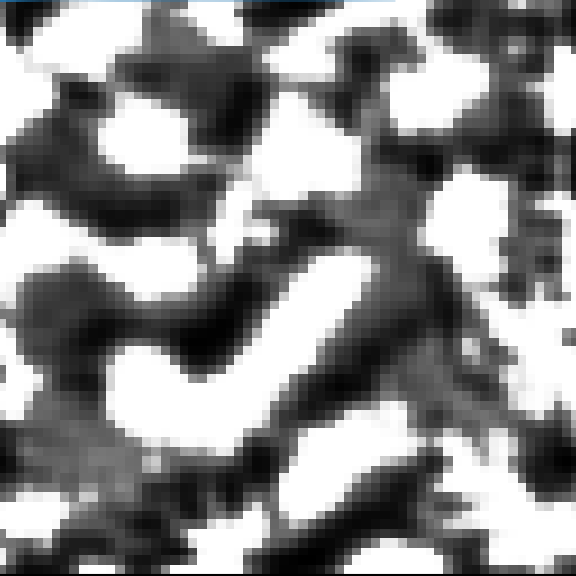} &
		\includegraphics[width=\wa\linewidth ]{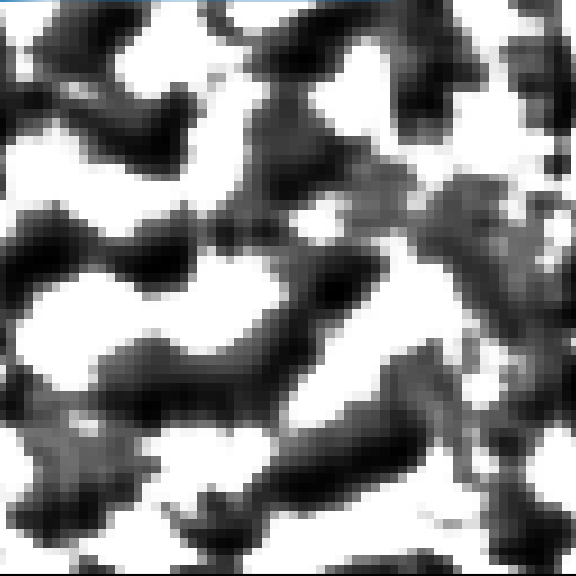} &	
		\includegraphics[width=\wa\linewidth ]{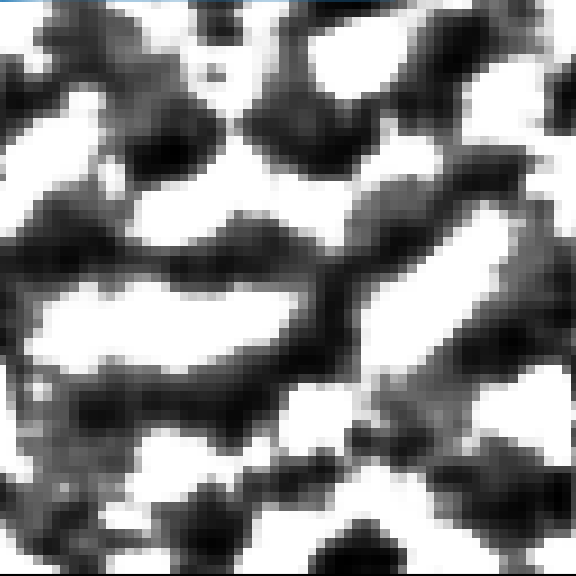} \\
		
		\raisebox{0.8em}{\small{(c)}} &
		\includegraphics[width=0.054\linewidth , cframe=ube 1pt]{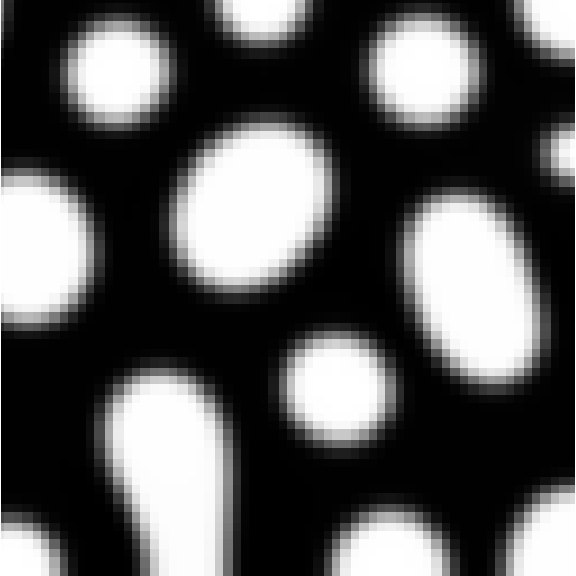} &
		\includegraphics[width=\wa\linewidth ]{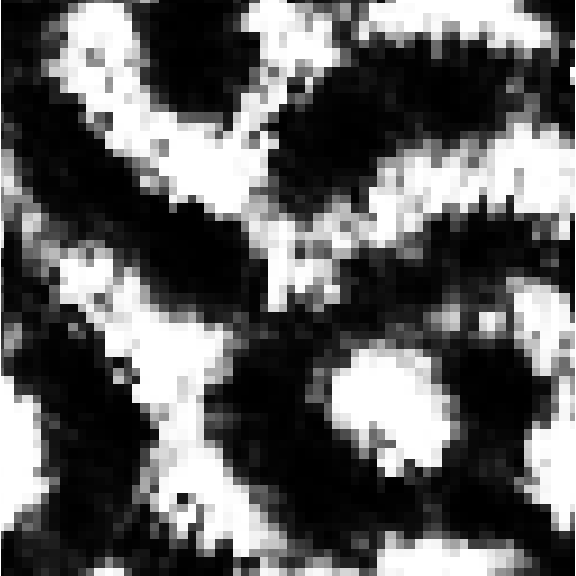} &
		\includegraphics[width=\wa\linewidth ]{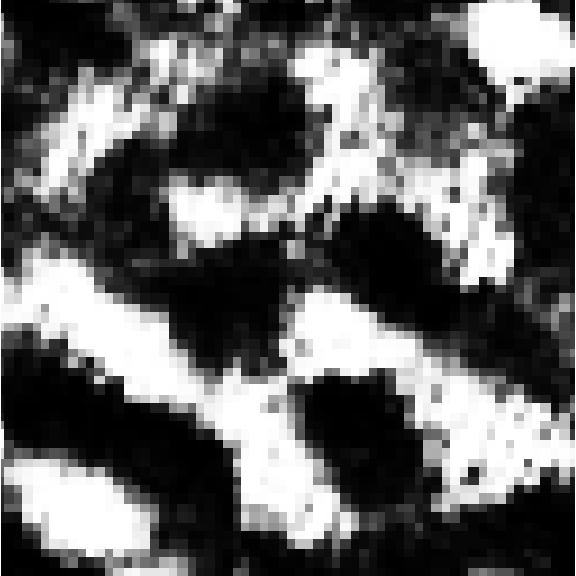} &
		\includegraphics[width=\wa\linewidth ]{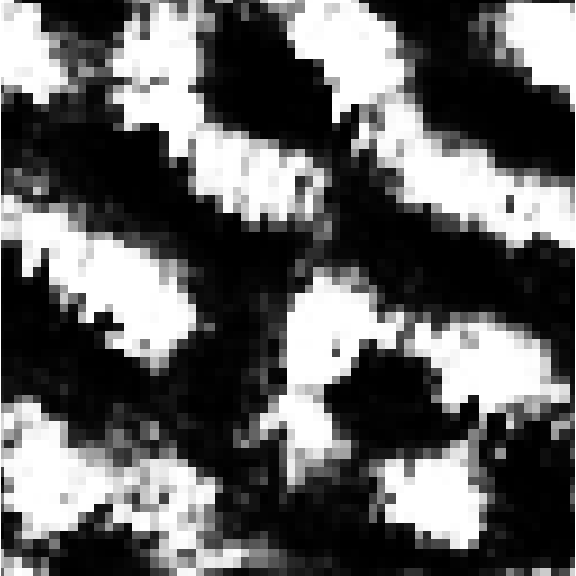} &
		\includegraphics[width=\wa\linewidth ]{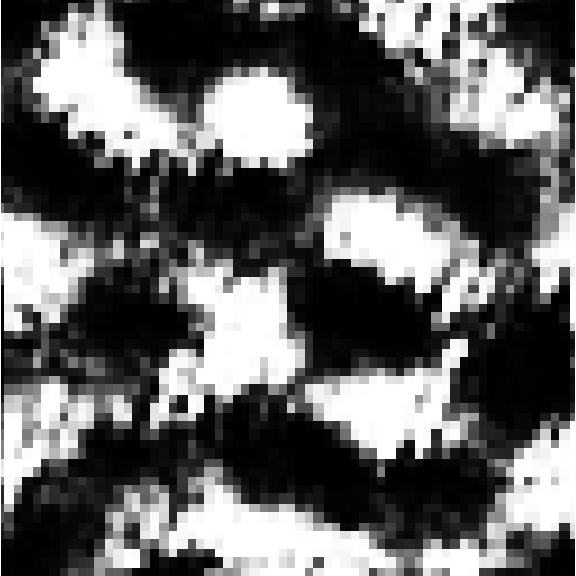} &
		\includegraphics[width=\wa\linewidth ]{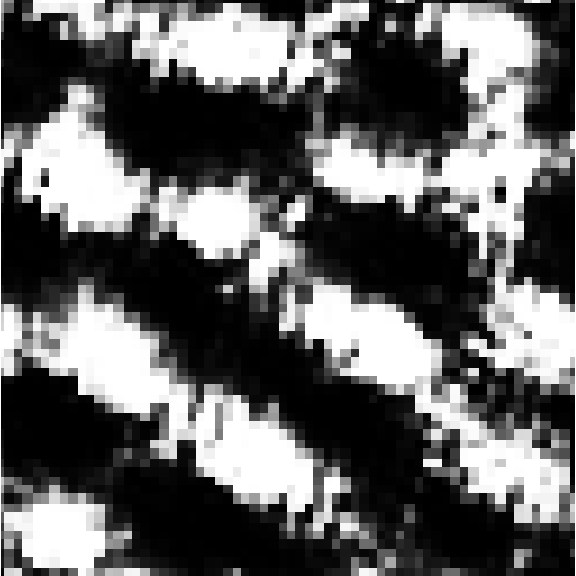} &
		\includegraphics[width=\wa\linewidth ]{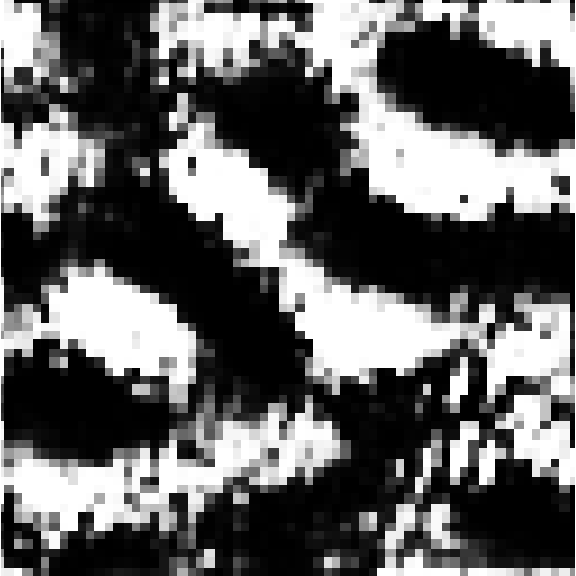} &
		\includegraphics[width=\wa\linewidth ]{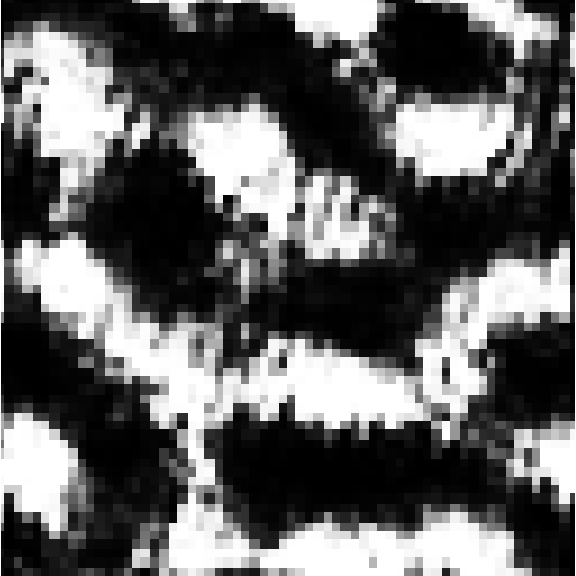} &
		\includegraphics[width=\wa\linewidth ]{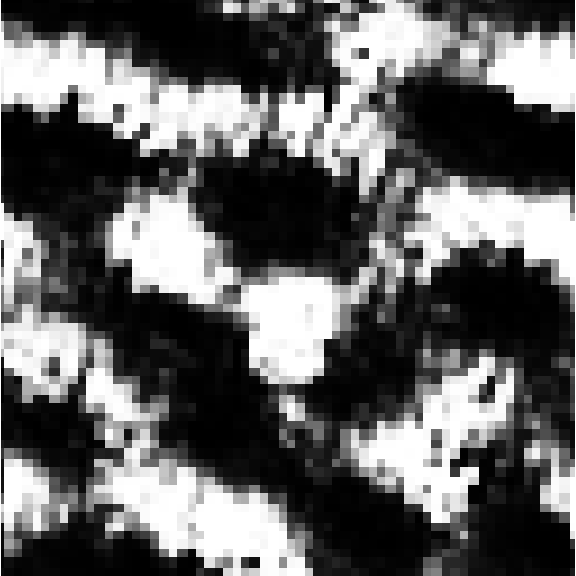} &
		\includegraphics[width=\wa\linewidth ]{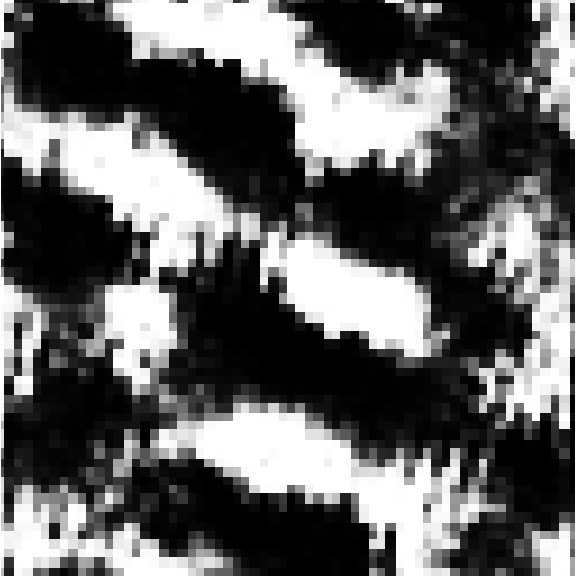} &
		\includegraphics[width=\wa\linewidth ]{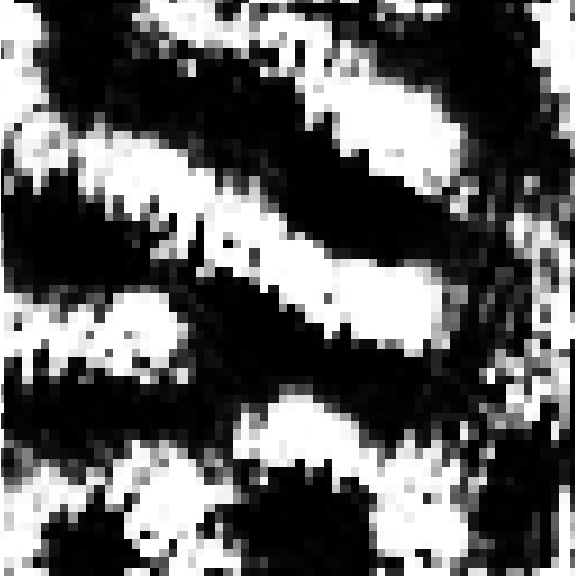} &
		\includegraphics[width=\wa\linewidth ]{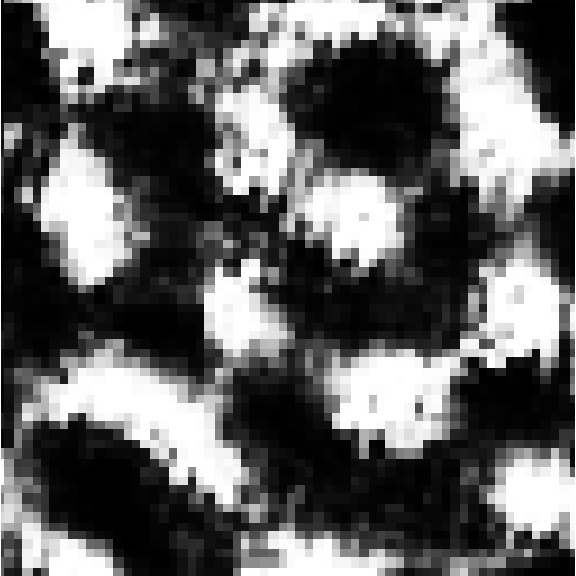} & 
		\includegraphics[width=\wa\linewidth ]{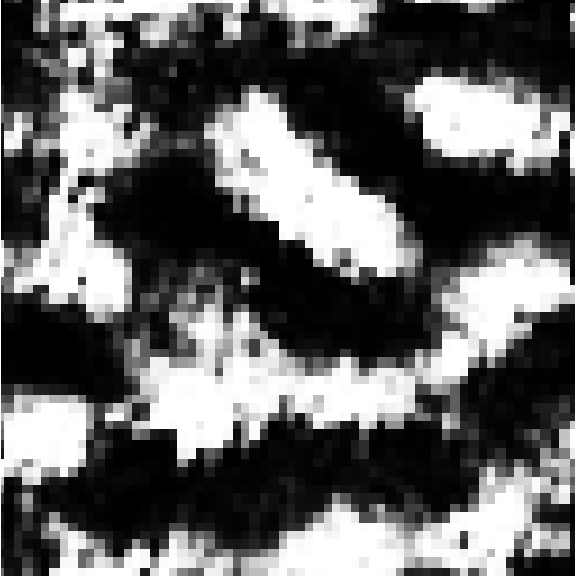} &
		\includegraphics[width=\wa\linewidth ]{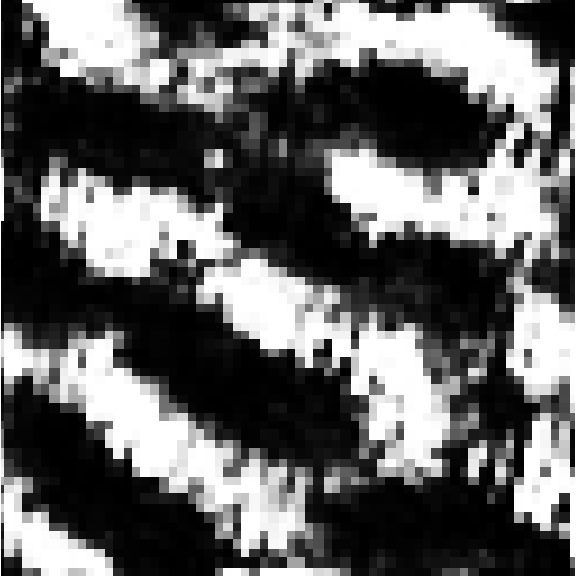} &
		\includegraphics[width=\wa\linewidth ]{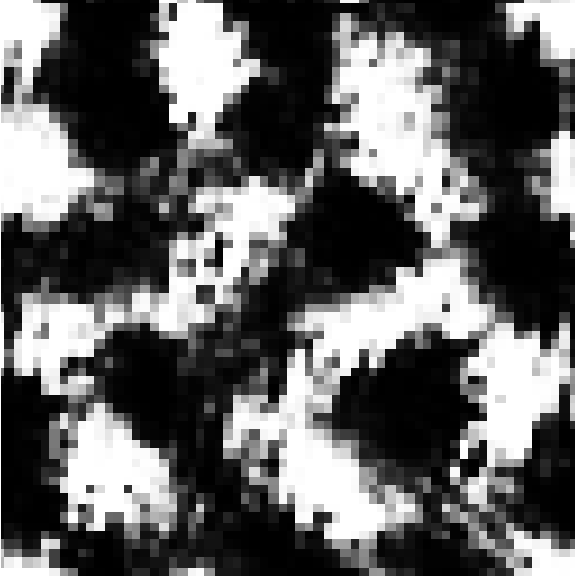} &	
		\includegraphics[width=\wa\linewidth ]{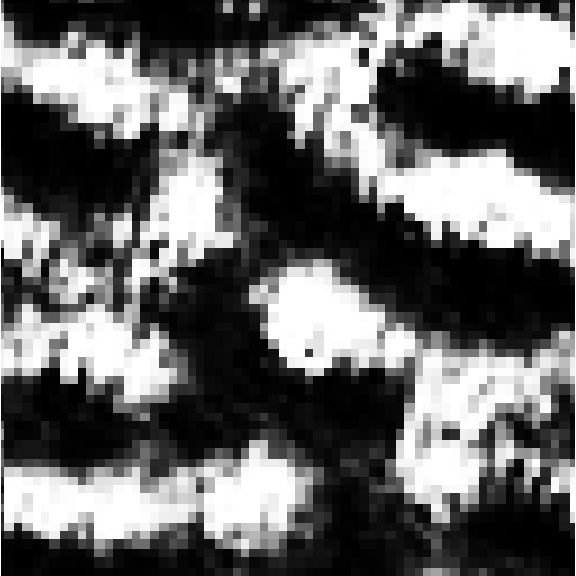} \\
		
	\end{tabular}
	\end{center}
	\caption{{Images generated by GIN models; with first image in each row being the real image used for calibration. (a,b) are trained over enite $p_2$ curve while model in (c) used only the initial portion of $p_2$ curve.}}
	\label{fig:invar1}
\end{figure}
\pgfkeys{/pgf/number format/.cd,1000 sep={\,}}
\usepgfplotslibrary{fillbetween}
 \begin{figure}[t]
     \begin{subfigure}[b]{0.34\textwidth}
          \centering
          \resizebox{\linewidth}{!}{
          \begin{tikzpicture}
           \tikzstyle{every node}=[font=\Large] 
          \begin{axis}[axis background/.style={fill=seabornback, fill opacity=1},
          ymin=0.25,
        ymax=0.45,
        ylabel=$p_2$ correlation,
        grid style={line width=.1pt, draw=white},
        major grid style={line width=.1pt,draw=white},
        minor tick num=1,
        grid=both,
        xlabel= Radial distance,
        legend cell align=left,
        legend style={at={(1,1)}}]
\addplot[ultra thick,color=ube] table[x=index, y=real, col sep=comma]{data/inv_p2_1_corrected.csv};
\addplot[thick,name path=max,color=amber] table[x=index, y=max, col sep=comma]{data/inv_p2_1_corrected.csv};
\addplot[thick,name path = min, color=amber] table[x=index, y=min, col sep=comma]{data/inv_p2_1_corrected.csv};
\addplot[color=amber, fill opacity=0.3] fill between[
    of = max and min];
  \legend{Original Image, , ,Generated Images}
 \end{axis} 
  \draw[shorten >=1mm,shorten <=1mm] (rel axis cs:0.20,1.20) node[right,draw,align=left]{
     \renewcommand{\arraystretch}{1.2}
     \begin{tabular}{cc}
$p_1$ & $0.413 \mp 0.018$ \\
$\frac{dp_2(r=0)}{dr}$  & $120\degree \mp 10\degree$ \\
\end{tabular}};
 \end{tikzpicture}} 
 \caption{\footnotesize{GIN trained on entire $p_2$ curve}}
\end{subfigure}
\begin{subfigure}[b]{0.32\textwidth}
          \centering
          \resizebox{\linewidth}{!}{
         \begin{tikzpicture}
          \tikzstyle{every node}=[font=\Large]
\begin{axis}[axis background/.style={fill=seabornback, fill opacity=1},
ymin=0.25,
ymax=0.45,
xlabel= Radial distance,
    grid style={line width=.1pt, draw=white},
    major grid style={line width=.1pt,draw=white},
    minor tick num=1,
     grid=both,
     legend cell align=left,
     legend style={at={(1,1)}}]
\addplot[ultra thick,color=ube] table[x=index, y=real, col sep=comma]{data/inv_p2_2_corrected.csv};
\addplot[thick,name path=max,color=amber] table[x=index, y=max, col sep=comma]{data/inv_p2_2_corrected.csv};
\addplot[thick,name path = min, color=amber] table[x=index, y=min, col sep=comma]{data/inv_p2_2_corrected.csv};
\addplot[color=amber, fill opacity=0.3] fill between[
    of = max and min];
    \legend{Original Image, , ,Generated Images}
 \end{axis}
  \draw[shorten >=1mm,shorten <=1mm] (rel axis cs:0.20,1.20) node[right,draw,align=left]{
     \renewcommand{\arraystretch}{1.2}
     \begin{tabular}{cc}
$p_1$ & $0.431 \mp 0.04$ \\
$\frac{dp_2(r=0)}{dr}$  & $110\degree \mp 10\degree$ \\
\end{tabular}};
 \end{tikzpicture}}
 \caption{\footnotesize{GIN trained on entire $p_2$ curve}}
\end{subfigure}
\begin{subfigure}[b]{0.31\textwidth}
          \centering
          \resizebox{\linewidth}{!}{
          \begin{tikzpicture}
           \tikzstyle{every node}=[font=\Large]
\begin{axis}[axis background/.style={fill=seabornback, fill opacity=1},
ymin=0.05,
ymax=0.35,
grid style={line width=.1pt, draw=white},
    major grid style={line width=.1pt,draw=white},
    minor tick num=1,
     grid=both,
     xlabel=Radial distance,
     legend cell align=left,
     legend style={at={(1,1)}}]
\addplot[ultra thick,color=ube] table[x=index, y=real, col sep=comma]{data/inv_p2_4.csv};
\addplot[thick,name path=max,color=amber] table[x=index, y=max, col sep=comma]{data/inv_p2_4.csv};
\addplot[thick,name path = min, color=amber] table[x=index, y=min, col sep=comma]{data/inv_p2_4.csv};
\addplot[color=amber, fill opacity=0.3,area legend] fill between[
    of = max and min];
    \legend{Original Image, , ,Generated Images}
 \end{axis}
     \draw[shorten >=1mm,shorten <=1mm] (rel axis cs:0.20,1.20) node[right,draw,align=left]{
     \renewcommand{\arraystretch}{1.2}
     \begin{tabular}{cc}
$p_1$ & $0.296 \mp 0.017$ \\
$\frac{dp_2(r=0)}{dr}$  & $120\degree \mp 0.3\degree$ \\
\end{tabular}};
 \end{tikzpicture}}
    \caption{\footnotesize{GIN trained on initial $p_2$ curve}}
     \end{subfigure}
     \caption{Comparisons of $2-$ point correlation curves between the images generated by GIN models and the target image.}
     \label{fig:invar2a}
 \end{figure}
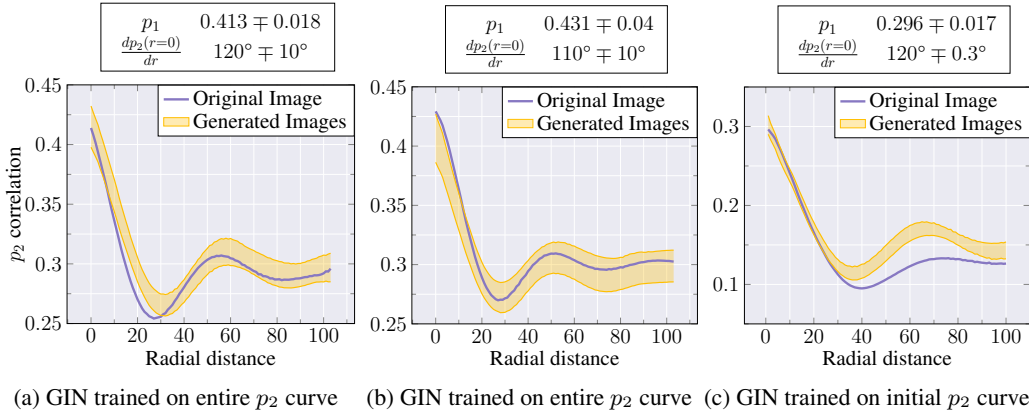
\subsection{Experiments on Hybrid (GAN+GIN) model}
\subsubsection{Architecture}
The hybrid (GAN+GIN) model combines both the adversarial discriminator and invariance checker in the training of the generator. The architectures of the generator (G) and the adversarial discriminator (D) are kept same as the WGAN-GP case (described in section~\ref{sec:wgan-arch}). The invariant checker is calibrated in a similar fashion as the GIN case. 

\subsubsection{Qualitative results}
The results for the GAN+GIN model are shown in Fig.~\ref{fig:hybridgraph}. The first image on the top right represents the original image and all the other images are synthetically generated by the hybrid model. The similarities are visually evident. Moreover, we also observe sufficient diversity in the simulated images. We further validate the efficacy of the model via the values of statistical properties as shown in Fig.~\ref{fig:hybridgraph}. We compare the value of $p_1$ and $p_2$ image descriptors of the synthetic images with the original image. The volume fraction of the original image is 0.436, and the distribution of volume fraction of synthetic images is in the range [0.418, 0.452]. Also, the distribution is symmetric around the mean and it follows a Gaussian distribution with mean at $\approx 0.436$. Similar values are observed from the $p_2$ curve shown in Fig.~\ref{fig:hybridgraph}(b). The $p_2$ curves of the synthetic images closely match the $p_2$ curve of the original image.  
Thus, the hybrid approach combines advantages from both the previous models, and is easily extendable to include higher order descriptors of the microstructures. 

\pgfkeys{/pgf/number format/.cd,1000 sep={\,}}
\usepgfplotslibrary{fillbetween}
 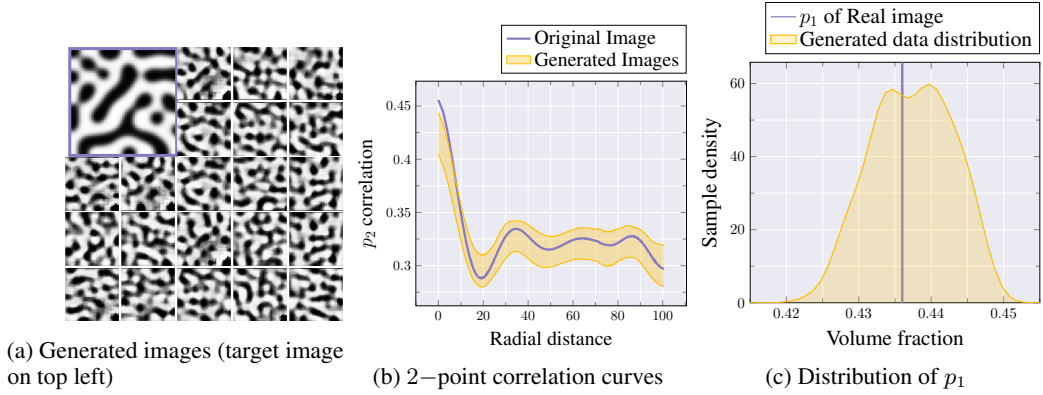
\begin{figure}[t]
 \begin{subfigure}[b]{0.32\textwidth}
 \centering
 \resizebox{\linewidth}{!}{
		\setlength{\tabcolsep}{0.2pt}
		\renewcommand{\arraystretch}{0.1}
		\begin{tabular}{ccccc}
			\multicolumn{2}{c}{\multirow{0}{0.108\linewidth}{\includegraphics[width=\linewidth,cframe=ube 0.4pt]{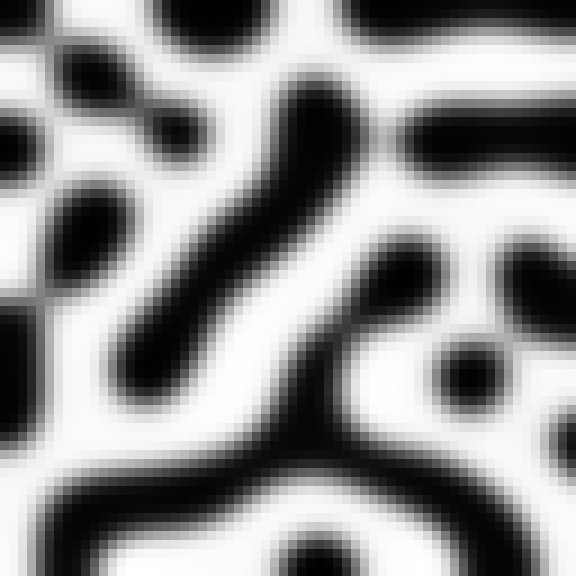}}} &
				   	& &   \\
					& &
			\includegraphics[width=\wa\linewidth ]{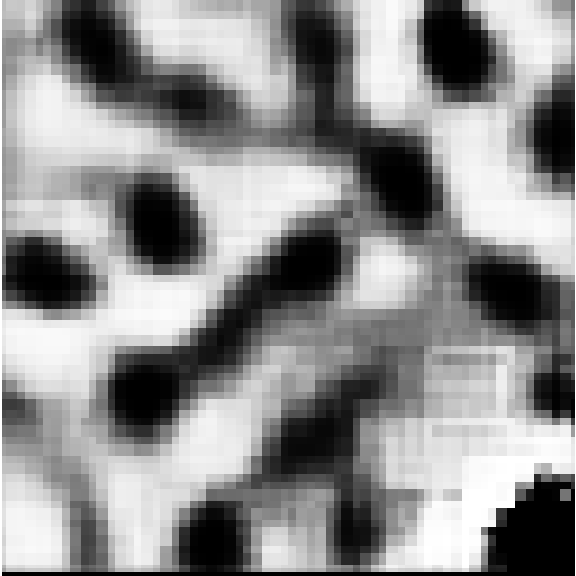} &
			\includegraphics[width=\wa\linewidth ]{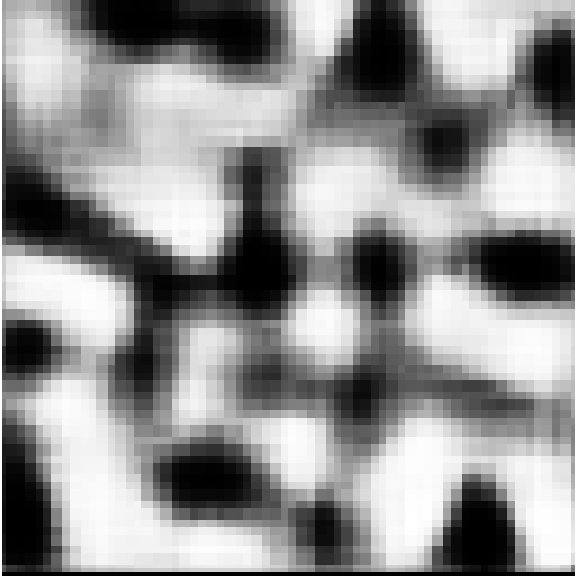} &
			\includegraphics[width=\wa\linewidth ]{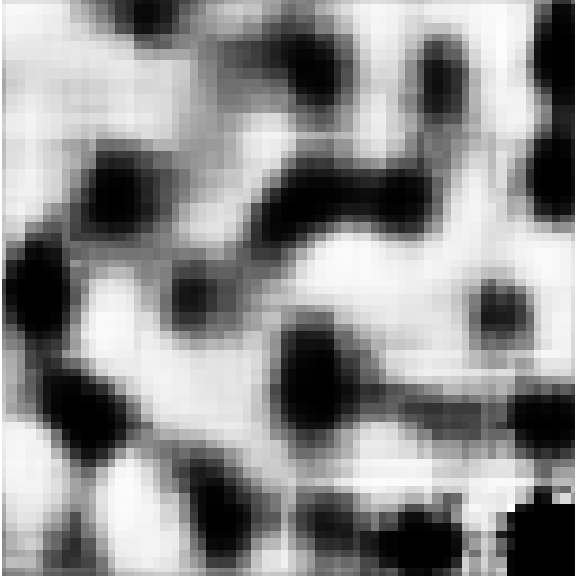}  \\
			 		& & 
			\includegraphics[width=\wa\linewidth ]{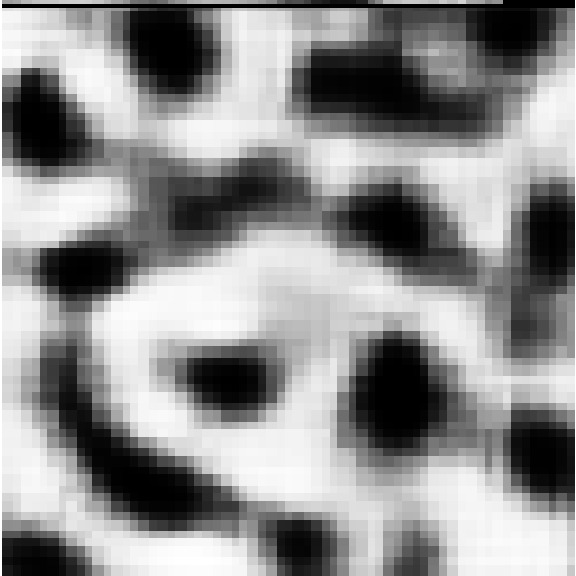} &
			\includegraphics[width=\wa\linewidth ]{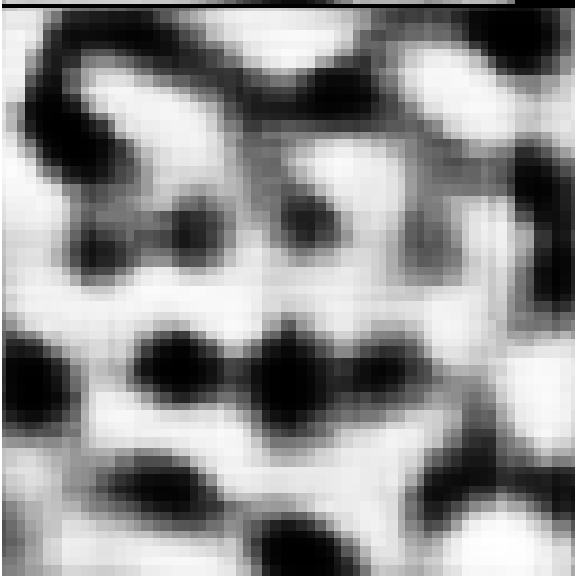} &
			\includegraphics[width=\wa\linewidth ]{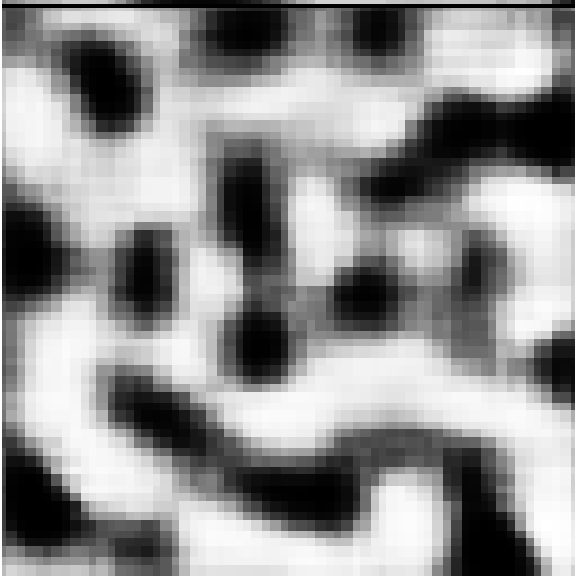} \\	
			\includegraphics[width=\wa\linewidth ]{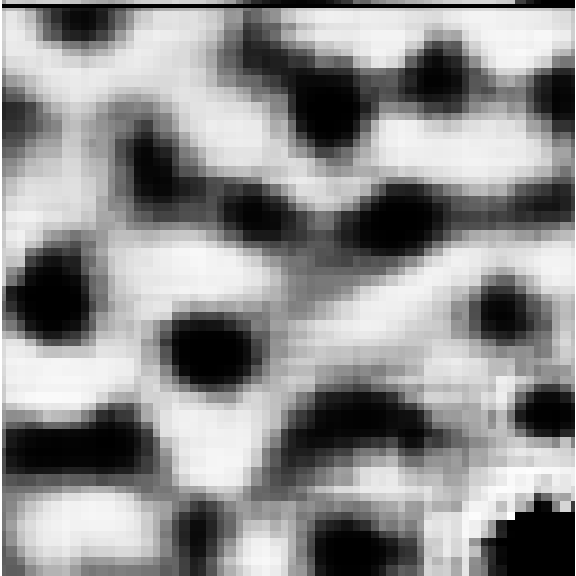} &
			\includegraphics[width=\wa\linewidth ]{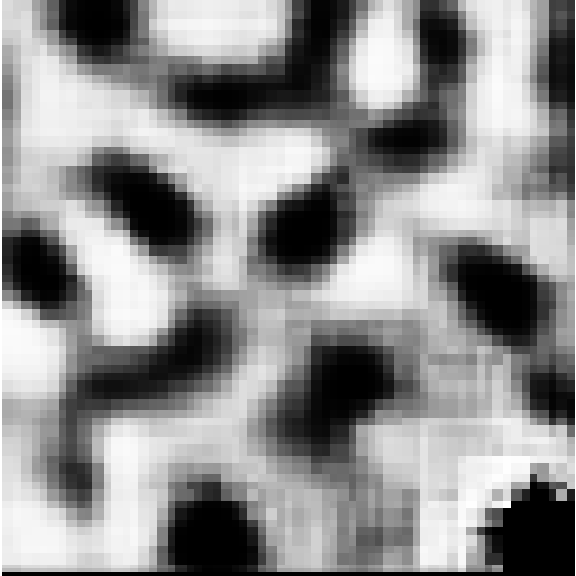} &
			\includegraphics[width=\wa\linewidth ]{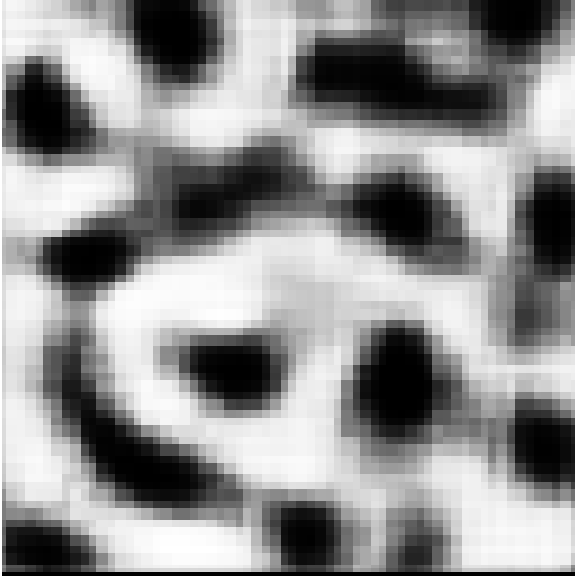} &
			\includegraphics[width=\wa\linewidth ]{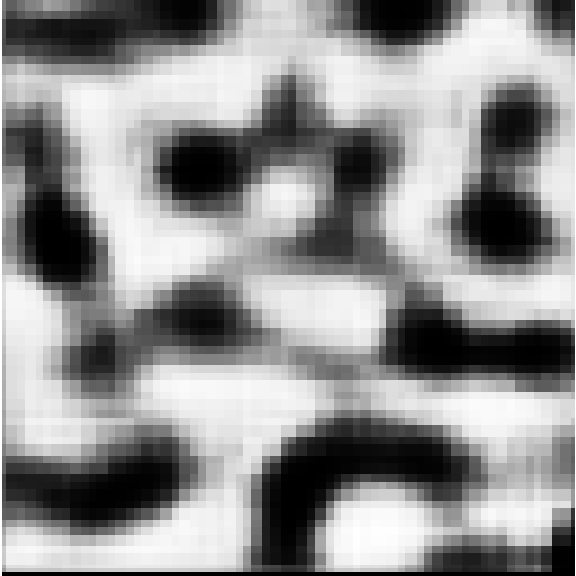} &
			\includegraphics[width=\wa\linewidth ]{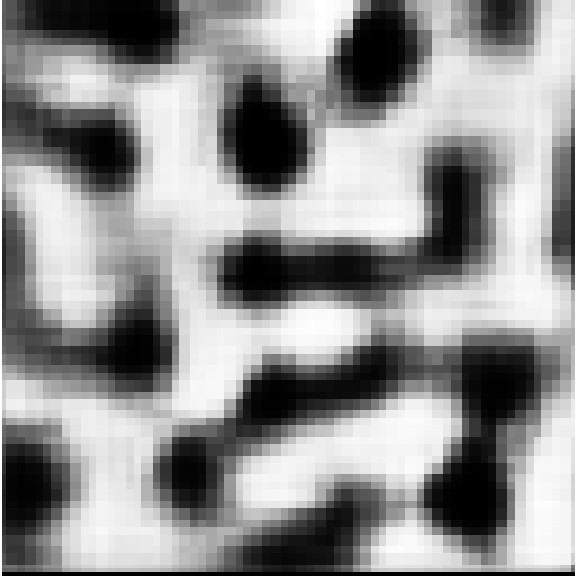}  \\ 
			\includegraphics[width=\wa\linewidth ]{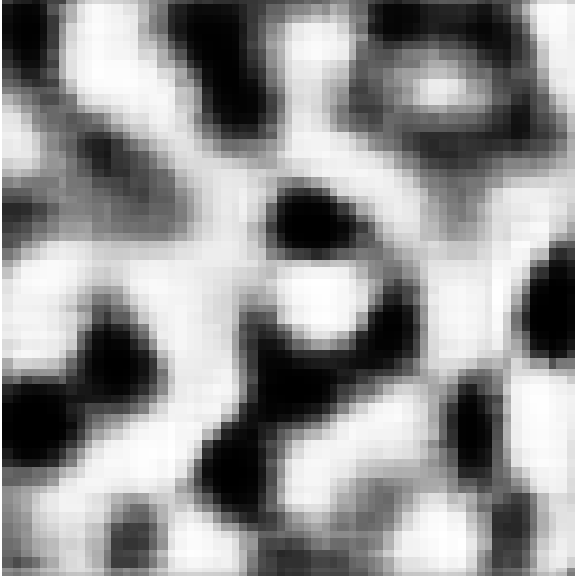} &
			\includegraphics[width=\wa\linewidth ]{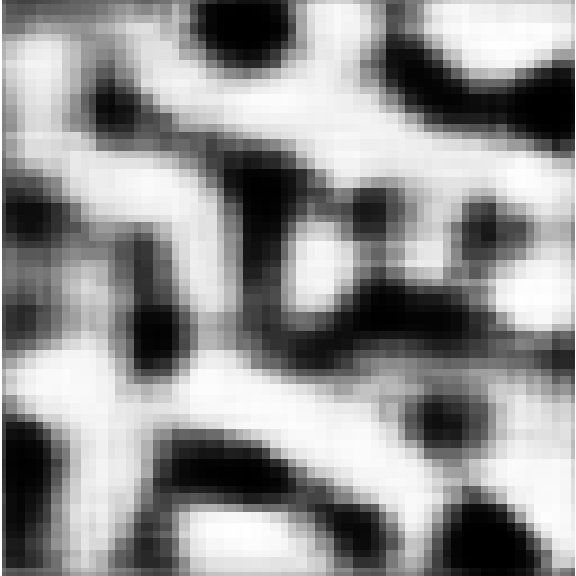} &
			\includegraphics[width=\wa\linewidth ]{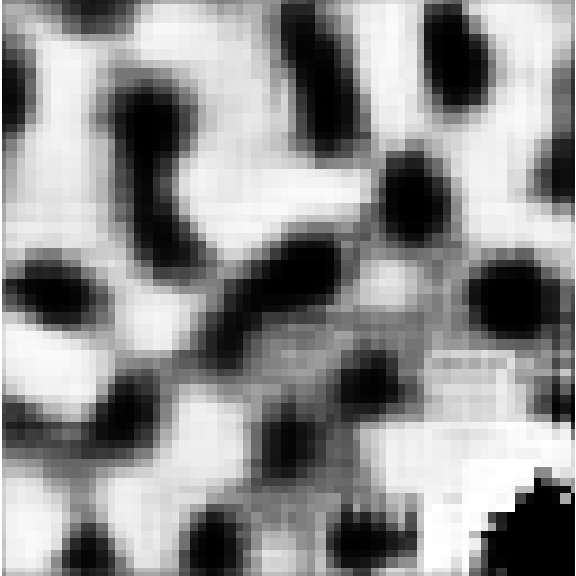} &
			\includegraphics[width=\wa\linewidth ]{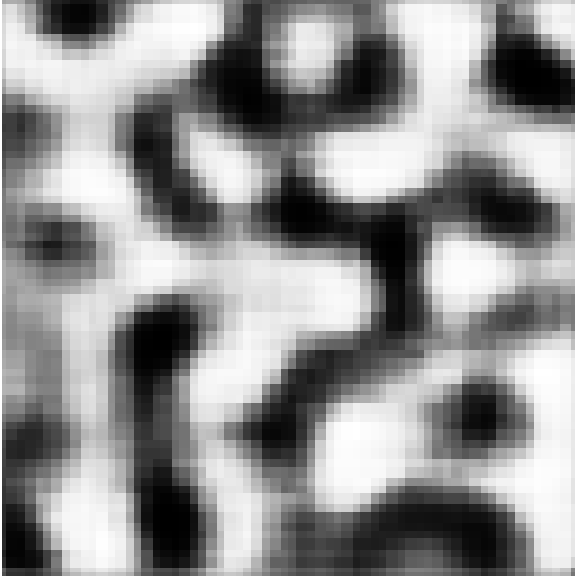} &
			\includegraphics[width=\wa\linewidth ]{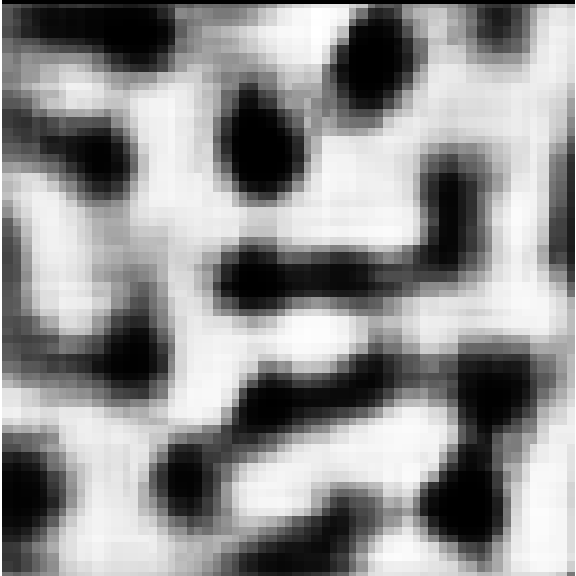}  \\ 
			\includegraphics[width=\wa\linewidth ]{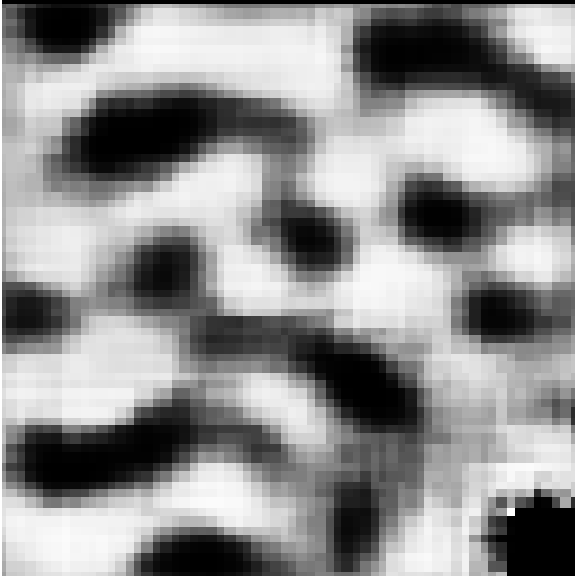} & 
			\includegraphics[width=\wa\linewidth ]{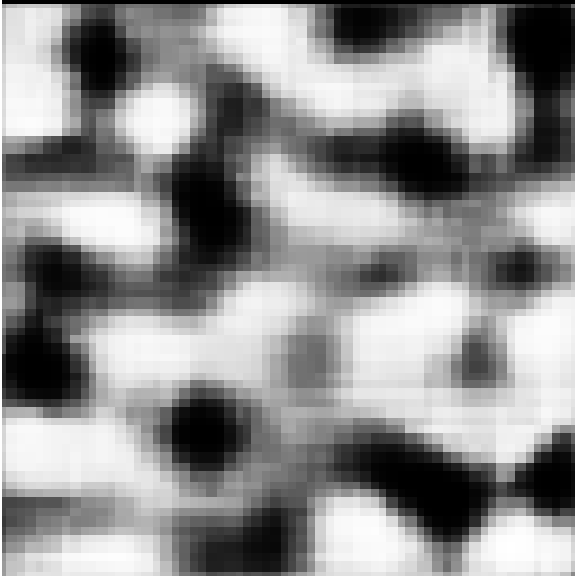} & 
			\includegraphics[width=\wa\linewidth ]{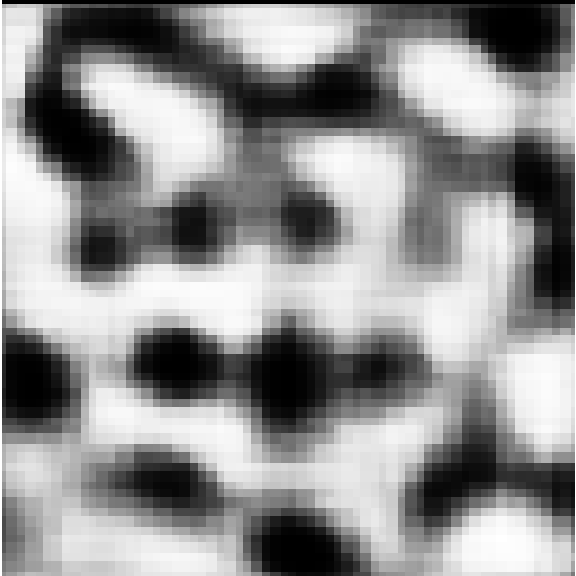} &
			\includegraphics[width=\wa\linewidth ]{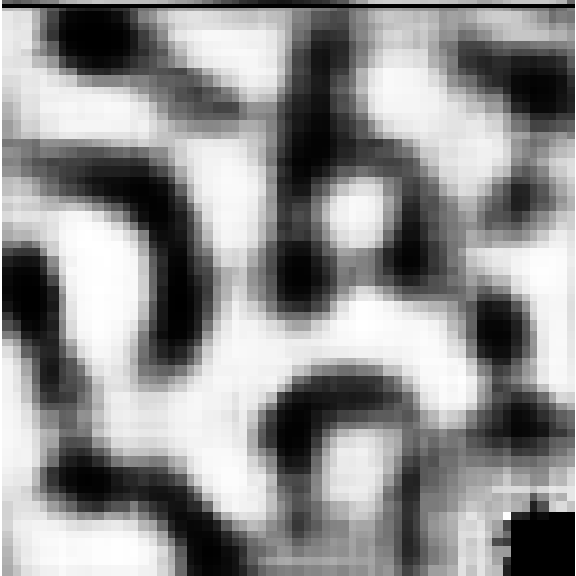} &
			\includegraphics[width=\wa\linewidth ]{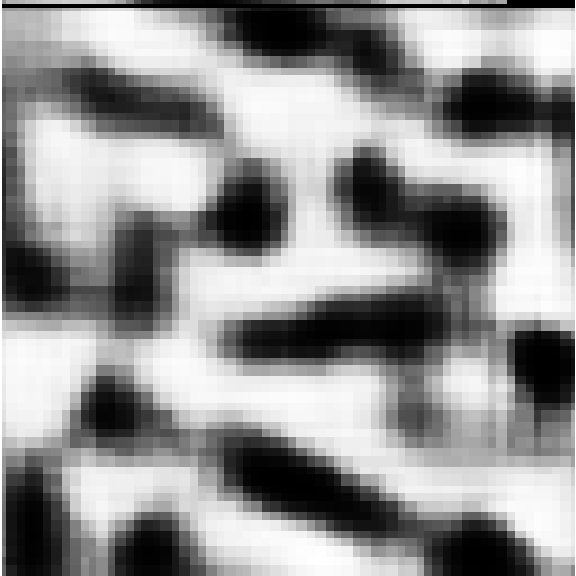}  \\	
		\end{tabular}}  
		\caption{\small{Generated images (target image on top left)}}
 \end{subfigure}
     \begin{subfigure}[b]{0.32\textwidth}
          \centering
          \resizebox{\linewidth}{!}{
          \begin{tikzpicture}
          \begin{axis}[axis background/.style={fill=seabornback, fill opacity=1},
        ylabel=\Large{$p_2$ correlation},
        grid style={line width=.1pt, draw=white},
        major grid style={line width=.1pt,draw=white},
        minor tick num=1,
        grid=both,
        xlabel= \Large{Radial distance},
        legend cell align=left,
        legend style={at={(1,1.25)}}]
\addplot[ultra thick,color=ube,x filter/.code={\pgfmathparse{\pgfmathresult*0.25}\pgfmathresult}] table[x=index, y=real, col sep=comma]{data/hybrid_p2.csv};
\addplot[thick,name path=max,color=amber,x filter/.code={\pgfmathparse{\pgfmathresult*0.25}\pgfmathresult}] table[x=index, y=max, col sep=comma]{data/hybrid_p2.csv};
\addplot[thick,name path = min, color=amber,x filter/.code={\pgfmathparse{\pgfmathresult*0.25}\pgfmathresult}] table[x=index, y=min, col sep=comma]{data/hybrid_p2.csv};
\addplot[color=amber, fill opacity=0.3] fill between[
    of = max and min];
  \legend{\Large{Original Image}, , ,\Large{Generated Images}}
 \end{axis}
 \end{tikzpicture}} 
 \caption{{$2-$point correlation curves}}
\end{subfigure}
\begin{subfigure}[b]{0.33\textwidth}
          \centering
         \resizebox{\linewidth}{!}{
\begin{tikzpicture}
 \begin{axis}[ymin=0,axis background/.style={fill=seabornback, fill opacity=1},
grid style={line width=.1pt, draw=white},
    major grid style={line width=.1pt,draw=white},
    minor tick num=1,
     grid=both,
     xlabel=\Large{Volume fraction},
     xmin=0.415,
     xmax=0.455,
     ylabel=\Large{Sample density},
     legend cell align=left,
     legend style={at={(1,1.25)}},ylabel style={at={(2ex,0.5)}}]
     \draw [ultra thick, color=ube]({axis cs:0.436,0}|-{rel axis cs:0,0}) -- ({axis cs:0.436,0}|-{rel axis cs:0,1});
\addlegendimage{color=ube,thick}
\addlegendentry{\Large{$p_1$ of Real image}}
\addplot[thick,color=amber,fill, fill opacity=0.2,area legend] table[x=x, y=kde, col sep=comma]{data/hybrid_p1.csv}; \addlegendentry{\Large{Generated data distribution}}
\end{axis}
\end{tikzpicture}}
 \caption{\small{Distribution of $p_1$}}
\end{subfigure}
\caption{Comparisons of volume fraction ($p_1$) distribution and $2-$ point correlation curves between the images generated by hybrid GAN and the target image.}
     \label{fig:hybridgraph}
 \end{figure}
 

{{\small
\bibliographystyle{plain}
\bibliography{./ref}
}}
\end{document}